\documentclass[aps,pra,floatfix,superscriptaddress,twocolumn]{revtex4-2}

\usepackage{graphicx,psfrag,color,bbm}
\usepackage{amsmath,amssymb}
\usepackage{mathtools}
\usepackage{hyperref}
\usepackage{bm}
\newcommand{\ket}[1]{|#1 \rangle}
\newcommand{\bra}[1]{\langle  #1 | }

\newcommand{\braket}[2]{\langle  #1 | #2 \rangle}

\newcommand{\average}[1]{\langle {#1} \rangle}

\newcommand{\E}{\mathrm{e}}
\newcommand{\D}{\mathrm{d}}

\newcommand{\tr}{\mathrm{Tr}}

\newcommand{\Psiu}{\Psi^{(1)}}
\newcommand{\bPsiu}{\bar \Psi^{(1)}}
\newcommand{\rhoz}{\rho^{(0)}}

\newcommand{\SB}{\Delta S_B}
\renewcommand{\SS}{\Delta S_S}
\newcommand{\tf}{t_{\mathrm f} }

\newcommand{\rf}{r_{\mathrm f} }
\newcommand{\rhof}{\rho^{(\mathrm f)} }
\newcommand{\rhoft}{\tilde \rho^{(\mathrm f)} }
\newcommand{\cS}{\mathcal{S}}

\newcommand{\Ht}{\mathcal{H}}
\newcommand{\rt}{\varrho}
\newcommand{\trt}{\tilde \rt}
\newcommand{\Ut}{U_t}
\newcommand{\Utt}{U_{\tilde t}}
\newcommand{\Pif}{\Pi_f}
\newcommand{\pif}{\pi_f}
\newcommand{\Nu}{\mathcal{J}}
\newcommand{\tpif}{\tilde \pi_f}
\newcommand{\rhob}{\rho_B}
\newcommand{\rhobt}{\tilde\rho_B}

\newcommand{\comm}[2]{\left[#1,#2\right]}
\newcommand{\al}{\alpha,\lambda}
\newcommand{\Hb}{H_B}
\newcommand{\Hba}{H_{B_\alpha}}
\newcommand{\Sba}{S_{B_\alpha}}
\newcommand{\Sb}{S_{B}}
\newcommand{\Hs}{H}
\renewcommand{\bar}[1]{\mkern 1.5mu\overline{\mkern-1.5mu#1\mkern-1.5mu}\mkern 1.5mu}
\newcommand{\HD}{H_{D}}
\newcommand{\HND}{H_{ND}}

\newcommand{\myO}{\bm{\varOmega}}

\newcommand{\p}[1]{\left({#1}\right)}
\newcommand{\pq}[1]{\left[{#1}\right]}
\newcommand{\pg}[1]{\left\{{#1}\right\}}

\newcommand{\dt}[1]{\frac{d #1 }{d t}}

\newcommand{\II}{\mathbbm{I}}
\newcommand{\ii}{\mathbbm{i}}

\usepackage{comment}

\begin{document}

\title{A quantum fluctuation theorem for dissipative processes}

\author{Gabriele De Chiara}
\affiliation{Centre  for  Theoretical  Atomic,  Molecular  and  Optical  Physics, Queen's  University  Belfast,  Belfast  BT7 1NN,  United  Kingdom}
\affiliation{Laboratoire Charles Coulomb (L2C), UMR 5221 CNRS-Universit\'e de Montpellier, F-34095 Montpellier, France}
\author{Alberto Imparato}
\affiliation{Department of Physics and Astronomy, Aarhus University, Aarhus 8000, Denmark}
\affiliation{Centre  for  Theoretical  Atomic,  Molecular  and  Optical  Physics, Queen's  University  Belfast,  Belfast  BT7 1NN,  United  Kingdom}

\begin{abstract}
We present a general quantum fluctuation theorem  for the entropy production of an open quantum system coupled to multiple environments, not necessarily at equilibrium.
 Such a general theorem, when restricted to the weak-coupling and Markovian regime, holds  for both local and global master equations, corroborating the thermodynamic consistency of  local quantum master equations.
The theorem is genuinely quantum, as it can be expressed in terms of conservation of a Hermitian operator, describing the dynamics of the system state operator and of the entropy change in the baths.
The integral fluctuation theorem follows from the properties of such an operator. 
 Furthermore, it is also valid when the system is described by a time-dependent Hamiltonian. As such, the quantum Jarzynski equality is a particular case of the general result presented here. Moreover, our result can be extended to nonthermal baths, as long as microreversibility is preserved.
We  present some numerical examples to showcase
the exact results previously obtained.
We finally generalize the fluctuation theorem to the case where the interaction between the system and the bath is explicitly taken into account. We show that the fluctuation theorem amounts to a relation between time-reversed dynamics of the global density matrix and  a two-time correlation function along the forward dynamics involving the baths’ entropy alone. 
\end{abstract}
\maketitle

\section{ Introduction} 

In classical stochastic thermodynamics the physics of work and heat fluctuations in  small, out-of-equilibrium  systems is 
now well understood in the framework of  fluctuation theorems (FTs) extending the second law of thermodynamics to the microscopic realm \cite{Jarzynski97,Kurchan98, Crooks99,Crooks00,Seifert05a,Esposito2010a,Seifert2012}.
Their importance originates from their generality, relying on very few assumptions, and from providing connections between equilibrium quantities and fluctuations of entropy, heat and work.
While the proof of the FT in the classical regime often relies on the stochastic trajectories that a system performs in its phase space while interacting with the external environment \cite{Seifert2012}, other approaches based on the symmetries of the classical master equation \cite{Imparato7a} or of the Fokker-Planck equation have been proposed \cite{Fogedby12,Fogedby14}.

In the quantum realm, several approaches have been put forward to generalise the FT and the impossibility of monitoring a quantum system without disturbance has generated a long debate (see, for example, the reviews~\cite{EspositoRMP2009,CampisiRMP2011,HanggiNP2015}). 
Several approaches to the quantum FT employ the formalism of quantum trajectories or quantum Monte Carlo (QMC)~\cite{DerezinskiJSTAT2008,LiuPRE2016,GarrahanPRL2010,HorowitzPRE2012,SubasiPRE2012,HorowitzNJP2013,LeggioPRA2013,ManzanoPRE2015,ManzanoPRX2018}. In this context, thermodynamic quantities, e.g. entropy and energy, can be sampled along quantum trajectories generated by a continuous evolution and random quantum jumps. Interestingly, the back-action from the observation of the jump, e.g. through the emission of a photon, gives rise to a genuinely quantum energy contribution, dubbed ``quantum heat''~\cite{Elouard2017,ElouardNJP2017,AuffevesSciPost2021}. Very often, the so-called two-point measurement scheme, where the system is observed at the start and end of the protocol, is used~\cite{KwonPRX2019}, potentially destroying  useful quantum coherences. Alternative approaches exist~\cite{ChetriteJSTAT2012,ManzanoPRL2019,MicadeiPRL2020,ManzanoPRL2021}. One can also prove the FT by diagonalizing  the instantaneous density matrix, thus generating a quantum counterpart of the classical trajectories \cite{Esposito2006}. 
A fully coherent quantum FT in the framework of quantum resource theory has also been proposed \cite{AbergPRX2018} and recently experimentally verified~\cite{Micadei2020}.

In this paper, we first present a general and unified approach that naturally extends the classical FT to open quantum systems and is only based on the quantum Lindblad master equation. 
We prove the FT by introducing an auxiliary quantum master equation and a Hermitian operator that accounts for the time evolution of both the system's state and the bath entropy.
Our formalism, based on the change of entropy in the baths and in the system, is general: it is valid at all times and not just at steady state; it is valid for both local and global quantum master equations (for which a heated debate has arisen in recent years), for an arbitrary number of baths,  and for time-dependent system Hamiltonians. 

We then consider the fate of the FT when the interaction between the system of interest and its environment is explicitly taken into account, and the total dynamics for the combined system is unitary.
We show that the FT in this case can be expressed in terms of a two-time correlation function involving the baths' entropy alone. Such a correlation function turns out to be equal to the transition probability of the system along the time-reversed dynamics. 
This result confirms the intimate connection between the FT and the lack of symmetry between the forward and backward dynamics.

Moreover, we  validate the results obtained in the first part of the paper: starting from the unitary dynamics we rederive the auxiliary master equation for the system in the same limit where the Lindblad master equation holds. We discuss explicitly the requirements for the local detailed balance to hold (or not), in connection with the interaction mechanisms between the system and the baths, and with the baths' properties. This analysis provides an operational approach to evaluate the entropy change in the bath and thus to test experimentally or numerically the proposed FT.

\begin{figure}[t]
\begin{center}
\includegraphics[width=0.8\columnwidth]{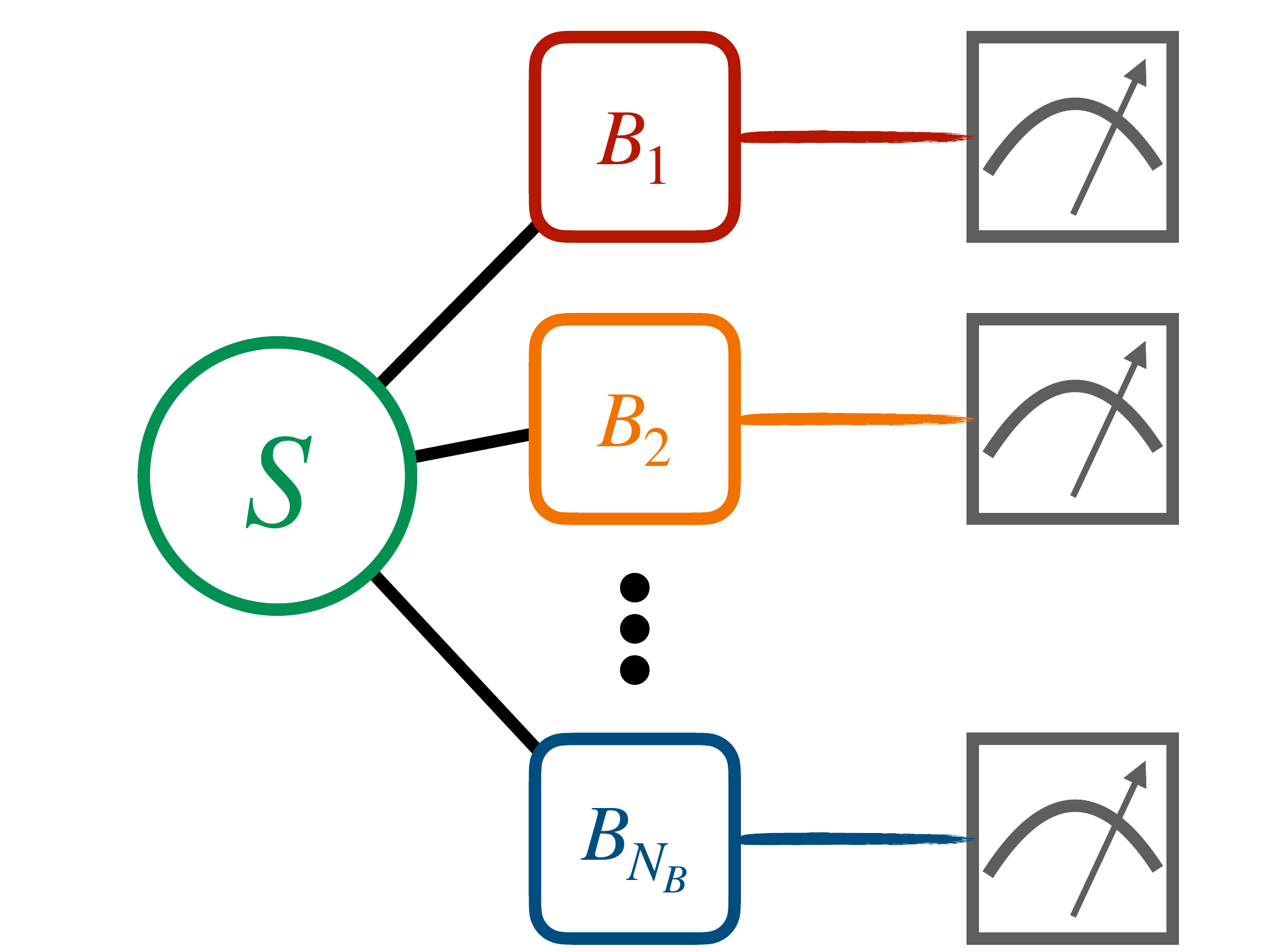}
\caption{Schematics of the setup: a system is interacting with $N_B$ baths which are continuously monitored by their respective measurement apparatus.}
\label{fig:setup}
\end{center}
\end{figure}

Our quantum FT goes beyond the two-point measurement scheme: it requires an initial projection of the system in an arbitrary basis to estimate the initial system's entropy along a specific trajectory. The initial projection may also occur in the eigenbasis of the initial density matrix, thus preventing any measurement back-action. It also requires the continuous monitoring of the baths but not the final system's projection or the continuous measurement of the work done on the system for time-dependent Hamiltonians (see Fig.~\ref{fig:setup}). As such, its experimental implementation may be easier than schemes based on two-point measurements.

Moreover, a dissipative quantum Jarzynski equality (JE) \cite{CrooksJSTAT2008,CampisiPRL2009,AlbashPRE2013,SilaevPRE2014} can be derived in operator form, and the result of the classical stochastic thermodynamics can be immediately recovered for diagonal density matrices.
Our approach is unifying as it provides a proof of the quantum FT for relevant physical situations, namely, manipulated systems in contact with multiple heat baths, and allows a full description of the system's thermodynamics

\section{The fluctuation theorem} 
\label{fluct:theo:ME}

We assume the evolution of a $d$-level system weakly coupled to $N_b$ environments to follow the Gorini-Kossakowski-Sudarshan-Lindblad (GKSL) master equation (ME)\cite{Davies74,Gorini76,Lindblad76} ($\hbar=1$):
\begin{equation}
\dt{ \rho(t)}=-\ii \pq{H(t),\rho(t)}+\sum_{\alpha=1}^{N_b} D_\alpha[\rho(t)],
\label{me:eq}
\end{equation}  
for its density matrix $\rho(t)$ with dissipators 
\begin{equation}
D_\alpha[\rho]=\sum_\lambda \gamma_{\alpha,\lambda}\left(L_\lambda \rho  L_\lambda^\dagger -\frac 1 2 \{ L_\lambda^\dagger   L_\lambda,\rho \}\right),
\label{diss}
\end{equation} 
where $\alpha$ labels the environment and $L_\lambda=\ket {j'} \bra j$ and $\lambda=\lambda( j\to j')$ denotes a transition between two states $\ket j$ and $\ket{ j'}$. In the following, in order to lighten the notation,  we will omit the initial and final states of such a transition.
The orthonormal basis $\{\ket j\}$, though arbitrary, is motivated by the physical environments. If the states $\ket j$ are the eigenstates of $H$
 the ME is dubbed ``global" or ``local" otherwise~\cite{HewgillPRR2021}.
We split an operator $X=X_D+X_{ND}$ into the sum of its diagonal and nondiagonal parts in the  basis $\{\ket j\}$ and set $X_{jj}=\bra j X \ket j$. 
We consider the general case where different baths can drive the same transition $\lambda$, if the dissipation rates (potentially time dependent) $\gamma_{\alpha,\lambda}\neq 0$.

In the classical case the FT reads \cite{Seifert2012}
\begin{equation}
\average{\E^{-\SB-\SS}}=1,\label{FTcl}
\end{equation}
where $\SS$ is the entropy change of the system along a given stochastic trajectory, 
and, if the detailed balance is fulfilled, the total entropy change in the baths due to the heat $Q_\alpha$ reads
 $\SB=-\sum_\alpha \beta_\alpha Q_\alpha$. In Eq.~(\ref{FTcl}) 
 the average runs over all the possible stochastic trajectories in the phase space, given the system dynamics and the time protocol $H(t)$.
We adopt the convention $Q_\alpha > 0$ when the heat flows from the bath into the system.
We will first derive a quantum FT without assuming the local detailed balance condition for the jump rates: such an assumption will be later introduced in the paper, and its consequences discussed.

The classical FT (\ref{FTcl}) requires the characterization of the bath entropy statistics. To this end, we will follow closely the approach discussed in Ref.~\cite{Imparato7a} for classical stochastic systems.
The ``jumps" between two states in the system  occur at the rates $\gamma_{j'j}$ because of the interaction with the baths:  
we thus introduce the elementary  current associated with the jump $\lambda$
\begin{equation}
\Delta s_{\alpha,j'j}= -\log(\gamma_{\alpha,jj'}/\gamma_{\alpha,j'j}).
\label{dsa}
\end{equation}

We introduce a total ``jump'' current $\cS$ given by the sum of the contributions (\ref{dsa}) for all the baths and for all the jumps up to time $t$. Such a quantity is akin to the entropy change $\SB$ in Eq.~(\ref{FTcl}): we will elaborate later on this connection. We would thus like to characterize the joint probability distribution $\Phi_j(\cS,t)$ of finding the system in the state $\ket j$ with a total jump current $\cS$  up to the time $t$. 
This can be done by introducing an extended quantum ME that takes into account the dynamic evolution of both the density operator and of the quantity $\cS$.
In practice we introduce a modified density matrix $\bar \rho(\cS,t)$, such that its diagonal elements describe the desired joint probability $\bar \rho_{jj}=\Phi_j(\cS,t)$. It is possible to show that the modified density matrix satisfies the modified ME:
\begin{eqnarray}
   && \partial_t \bar \rho_{j'j'}(\cS,t)=-\ii [H,\bar \rho]_{j'j'}+\nonumber \\
&&\, \sum_{\alpha,j}
    \left\{ \gamma_{\alpha,j'j} \left[\sum_{n=0}^\infty
    \frac{\left(-\Delta s_{\alpha,j'j}\right)^n}{n!}
    \frac{\partial^n \bar \rho_{jj}}{\partial \Delta S_B^n} \right]
    -\gamma_{\alpha,jj'}\bar \rho_{j'j'}\right\},\label{rhohD}\\
&&\partial_t \bar \rho_{lk}(\cS,t)=-\ii [H,\bar \rho]_{lk}+\sum_\alpha D_{ND,\alpha}[\bar \rho_{ND}]_{lk}, \, l\neq k, \label{rhohND}
\end{eqnarray} 
where $D_{ND,\alpha}[\cdot]$ is a superoperator such that $D_{ND}[\bar \rho_{ND}]$ has vanishing diagonal terms and  only couples nondiagonal terms of the density matrix in the chosen basis. A derivation of Eqs.~\eqref{rhohD}--\eqref{rhohND} based on the GKSL \eqref{me:eq} alone is discussed in Appendix~\ref{mod:app}.  
Such a derivation starts from  the fact that, if $\cS$ is the total jump current at time $t$, after the transition $j\to j'$ such a current reads $\cS+ \Delta s_{\alpha,j'j}$.
Next, we  introduce the generating function operator $\Psi(\xi,t)$ 
akin to the  generating function in classical physics, defined as
\begin{equation}
\Psi(\xi,t)=\int\, d \cS \, \bar \rho(\cS,t) \E^{-\xi \cS},
\label{psi:def}
\end{equation} 
and applying this integral transform to both sides of Eqs.~(\ref{rhohD})-(\ref{rhohND}), one obtains an equation for $\Psi(\xi,t)$
\begin{equation}
\partial_t\Psi(\xi,t)= \mathcal{L}_\xi[\Psi(\xi,t)],
\label{eq:Psi} 
\end{equation} 
where the superoperator $\mathcal{L}_\xi[\cdot]$ depends parametrically on $\xi$, and its full expression is given in Appendix \ref{mod:app}, see Eqs.~\eqref{PsiD:app}--\eqref{PsiND:app}.
We will see that the modified density matrix $\bar \rho(t)$,  its generating function $\Psi(t)$  and their modified MEs  are the essential ingredients to prove the quantum FT.

We  now discuss the physical initial condition for Eqs.~\eqref{rhohD}-\eqref{rhohND} and Eq.~\eqref{eq:Psi}. Let $\rhoz$ be the initial state at $t=0$. Since no current $\cS$ has yet been generated: 
$\bar \rho(\cS,t=0)=\rhoz \delta(\cS)$, where $\delta(x)$ is the Dirac-delta function.
Given the definition of $\Psi$, Eq.~\eqref{psi:def}, its initial condition becomes $\Psi(\xi,t=0)=\rhoz,\; \forall \xi$ and in
the following $\Psi(\xi,t|\, \rhoz)$ will indicate the solution to Eqs.~\eqref{eq:Psi} with this specific initial condition. The following equalities hold
\begin{eqnarray}
\average{ \E^{-\xi \cS}}_{\pi_{j},\rhoz}&=&\tr [ \pi_j \Psi(\xi,t| \rhoz)]=\Psi_{jj}(\xi,t| \rhoz),\label{Psi_jj} \\
\average{ \E^{-\xi \cS}}_{\rhoz}&=&\tr [ \Psi(\xi,t| \rhoz)],
\label{tr:Psi}
\end{eqnarray} 
where $\average{\cdots}_{\pi_j,\rhoz}$ in Eq.~(\ref{Psi_jj}) denotes the  expectation value constrained by the initial state $\rhoz$, the  state at time $t$  taken to be $\pi_j=\ket j\bra j$, and in Eq.~\eqref{tr:Psi} we summed over all possible final states.

In the following we will mostly be interested in the case $\xi=1$ to evaluate the expectation value of $\exp(-\cS)$. Letting $\Psiu(t)=\Psi(\xi=1,t)$, one finds
\begin{equation}
    \partial_t \Psiu(t) =-\ii [H,\Psiu(t)]+ \sum_\alpha D_\alpha^*[\Psiu(t)],
\label{eqPsidual}
\end{equation} 
 where $D_\alpha^*$ is the dual of the dissipator $D_\alpha$:
\begin{equation}
D_\alpha^*[\cdot]=\sum_\lambda \gamma_{\alpha,\lambda}\left(L_\lambda^\dagger  \cdot   L_\lambda -\frac 1 2 \{ L_\lambda^\dagger   L_\lambda,\cdot \}\right).
\label{diss:du}
\end{equation} 
The details of the derivation of Eq.~\eqref{eqPsidual} are given in Appendix~\ref{mod:app}.
Thus the time evolution of $\Psiu$, as given by Eq.~\eqref{eqPsidual},   has the same conservative part as in Eq.~\eqref{me:eq}, but the dissipative part (the dual of $D_\alpha$) is the same as the one found in the time evolution of an operator in the Heisenberg picture: this observation reflects the fact that the operator $\Psiu$ bears information on both the system state $\rho(t)$ and on the quantity $\exp(-\cS)$. As such, the operator $\Psiu$ is Hermitian at any time if it is Hermitian at $t=0$.
Modified MEs of the type \eqref{eq:Psi} emerge quite naturally through the large deviation approach used to study the long time limit of thermodynamic currents. Such an approach was first introduced in Ref.~\cite{Lebowitz1999}, and became later quite consolidated in the field of stochastic thermodynamics, see, e.g., Refs.~\cite{Imparato7a,Esposito2007, Chetrite2015}. A modified quantum ME was first introduced in \cite{GarrahanPRL2010} and later in \cite{Carollo2018} to study the long-time counting statistics in dissipative quantum systems. 
While the modified ME~\eqref{eqPsidual} above has been obtained by applying the integral transformation \eqref{psi:def} on the modified ME~\eqref{rhohD}--\eqref{rhohND}, in Sec.~\ref{uni:sec} we will provide an alternative derivation of Eq.~\eqref{eqPsidual}, based on the unitary evolution of a system explicitly interacting with a set of baths.

Let us now introduce the operator $ \bPsiu(t)$ 
\begin{equation}
\bPsiu(t)=\sum_{j_0} \Psiu(t| \pi_{j_0}).
\label{psi1Id}
\end{equation} 
Given that  $\bPsiu(t)$ is a linear combination of solutions of Eq.~\eqref{eq:Psi} with $\xi=1$, it is a solution itself, with initial condition 
$\bPsiu(0)=\sum_{j_0} \pi_{j_0}=\II$. Inspection of  Eq.~\eqref{eqPsidual}, suggests that  $\bPsiu(t)$ is a stationary solution at any time:
\begin{equation}
\bPsiu(t)=\bPsiu(t|\, \II)=\II, \quad \forall t\ge0.
\label{eq:bPsiu}
\end{equation} 
The last equality for $\bPsiu(t)$ is the first important result of the present paper.
It is the operatorial counterpart of the integral theorem (\ref{FTcl}), and it involves the Hermitian operators $\Psiu(t\,  |\, \pi_{j_0} )$ expressing the joint dynamics of the system density operator and of the total jump current $\cS$.
It does not depend on the choice of the initial basis. Remarkably by introducing an arbitrary basis  $\{\ket{b_0}\}$, and noticing  that Eq.~\eqref{eq:Psi} is linear, we can write the solution at time $t$, with the specific initial condition $\pi_{b_0}$ as
\begin{equation}
\Psiu(t\,  |\, \pi_{b_0})=\sum_{j_0,j_0'} \braket {j_0}{b_0}\braket{b_0}{j_0'} \Psiu(t\,  |\, \ket{j_0} \bra{j_0'}).
\label{eq:psiuk0}
\end{equation}
From Eq.~(\ref{eq:psiuk0}) one obtains $\sum_{b_0} \Psiu(t\,  |\, \pi_{b_0})= \sum_{j_0} \Psiu(t| \pi_{j_0})=\bPsiu(t)$, 
using $\sum_{b_0} \ket {b_0}\bra {b_0}=\II $.
 Furthermore Eqs.~\eqref{tr:Psi},\eqref{psi1Id}, and (\ref{eq:bPsiu}) imply $\sum_{j_0}\average{\exp(-\Delta S_B)}_{\pi_{j_0}}=d$.

We now present the mathematical statement of the integral FT which is the second main result of this paper,  and later on we discuss its physical significance. Such a FT reads:
\begin{eqnarray}
&&\tr [ \sum_{b_0} \rhoz_{b_0b_0} \Psiu(t\,  |\, \pi_{b_0}) \E^{-\log \rhoz_{b_0b_0}} \E^{\log \rhof}]=\nonumber \\
&&=\tr [  \bPsiu(t)  \rhof]=1\label{FTcoh},
\end{eqnarray} 
where we have used Eqs.~\eqref{psi1Id}-\eqref{eq:psiuk0} and introduced an arbitrary but normalised final state $\rhof$.
This feature of the FT was already noticed in Ref.~\cite{Seifert2012} for the classical case: Eq.~\eqref{FTcoh} holds for any normalised quantum final state $\rhof$, not necessarily the solution of Eq.~(\ref{me:eq}) at time $t$.

Eq.~\eqref{FTcoh} has the form of the expectation value of the Hermitian operator $\bPsiu$ over the final state of the system. We then  recall the physical interpretation of the operator $\Psiu(t|\, \rhoz)$: its $j$th diagonal element represents the   expectation value of $\exp(-\cS)$ constrained by the initial $\rhoz$ and final state $\pi_{j}$, Eq.~\eqref{Psi_jj}.
We make no specific assumption on the coherence in the initial and the final states  $\rho^{(0)}$  and $\rhof$: they can both exhibit coherence in the jump-operators' basis $\{\ket j\}$ as well as in the energy eigenbasis.
Let $\{\ket {k_0}\}$ and $\{\ket {\rf}\}$ be the  eigenbases of $\rhoz$ and $\rhof$, respectively.
 Changing basis, from  $\{\ket j\}$ 
to  $\{\ket{\rf}\}$ 
and recalling  definition \eqref{psi:def}, we can rewrite Eq.~\eqref{FTcoh} as 
\begin{eqnarray}
&&1=\sum_{\rf,j_0}\average{\E^{-\cS}}_{\pi_{\rf},\pi_{j0}} p_{j_0}^{(0)}  \E^{-\log \rhoz_{j_0j_0}} \E^{\log \rhof_{\rf \rf}}= \label{FTcoh2}\\
&=&\sum_{\rf,k_0}\int \D \cS \, \E^{-\cS+\log \rhof_{\rf \rf}-\log \rhoz_{k_0k_0}} \Phi_{\rf}(\cS,t|  \pi_{k_0}) p_{k_0}^{(0)}. \nonumber
\end{eqnarray} 
We notice that the last expression for the FT has the form of the classical counterpart  Eq.~\eqref{FTcl}, with an average over the joint probability distribution $\Phi_{\rf}(\cS)$.
Moreover, Eqs.~\eqref{FTcoh}-\eqref{FTcoh2} exhibit the structure of a double trace, one over the initial state $\rhoz$ and one over the final one $\rhof$, and as such their values do not depend on the chosen basis. Notice that performing the initial projection of $\rhoz$ in its  eigenbasis preserves its quantum coherence in other bases, e.g. in the energy eigenbasis. 
Using  Jensen's inequality, one recovers the second law: 
\begin{equation}
\average{\cS}-(\tr[\rhof \log \rhof]-\tr [\rhoz\log \rhoz])\ge 0.
\label{sec:law}
\end{equation}

We can  now establish the relation between the dynamics of the system, as expressed by Eqs.~(\ref{me:eq})-(\ref{diss}) and the thermodynamics significance of our results in Eqs.~(\ref{eq:bPsiu}), (\ref{FTcoh}), (\ref{FTcoh2}).
Such a relation follows immediately if the dissipation rates $\gamma_{\alpha,\lambda}$ are taken to obey the local detailed balance condition (LDBC):
${\gamma_{\alpha,\lambda} (\omega_{\lambda})}/{\gamma_{\alpha,\lambda} (-\omega_{\lambda})}=\exp(-\beta_\alpha \omega_{\lambda})$,
where 
$\omega_\lambda=\omega_{j'j}=H_{j'j'}-H_{jj}=\Delta H_{D,\lambda},$
and $\beta_\alpha$ is the inverse temperature of the bath $\alpha$.
If the LDBC holds then  
$\Delta s_{\alpha,j'j}$ in Eq.~(\ref{dsa}) can be immediately interpreted as the entropy change in the bath, given that   $-\omega_{j'j}=\beta_\alpha^{-1} \log(\gamma_{\alpha,j'j}/\gamma_{\alpha,jj'})$ is heat flowing into a bath as a consequence of a jump.
Thus the jump current $\cS$ entering Eqs.~(\ref{rhohD})-(\ref{psi:def}) can be identified, in this case, with the total change of entropy in the baths $\cS=\SB$.
While the LDBC is a standard assumption that allows to relate the dynamics to the thermodynamics in systems in contact with thermal baths, both in the classical and in the quantum regime \cite{Lebowitz1999, Seifert05a,ManzanoPRX2018}, our proof of the FT does not rely on it.
When the LDBC holds, evaluating the energy change  $-\omega_{j'j}$ in the bath is sufficient to evaluate the entropy change in it according to Eq.~(\ref{dsa}).  In case the LDBC does not hold, the rates $\gamma_{\alpha,j'j}$ and their relation to the $\omega_{j'j}$ need be estimated \emph{a priori}.

Assuming the LDBC, since the heat flowing into a bath as a consequence of a jump is $-\omega_{j'j}$, involving only the diagonal part of the system Hamiltonian, 
we have $\average{\cS}=\average{\SB}=-\sum_{\alpha}\beta_\alpha Q_{D,\alpha}$ and
we can rewrite Eq.~\eqref{sec:law} as
\begin{equation}
\SS=-(\tr[\rhof \log \rhof]-\tr [\rhoz\log \rhoz])\ge \sum_{\alpha}\beta_\alpha Q_{D,\alpha}.
\label{sec:lawa}
\end{equation} 
This is in accordance with the findings of Ref.~\cite{HewgillPRR2021}, where it was found that, for local MEs,  only the diagonal part of the heat currents, defined as $\dot  Q_{D,\alpha}=\tr\pg{ \rho D^*_\alpha[H_D]}$, enters the differential version of the second law: $d_t S_{S} \ge \sum_\alpha  \beta_\alpha \dot Q_{D,\alpha} $ and flows to the environments as shown in Appendix \ref{curr:app} in the Born-Markov approximation. Notice that both the diagonal and nondiagonal heat contributions enter the first law already at operatorial level as $d_t H=\sum_\alpha D^*_\alpha[H_D+H_{ND}]$  \cite{HewgillPRR2021}. However the nondiagonal part of the Hamiltonian, specifically the quantity $\dot  Q_{ND,\alpha}=\tr\pg{ \rho D^*_\alpha[H_{ND}]}$,  enters the energy balance of the interaction Hamiltonian, once a microscopic model made of the system, the baths and their interaction mechanism is considered as detailed in  Appendix \ref{curr:app}. 
While, within the collisional model framework, this corresponds to the work done when switching on and off the interaction with the environment, how this current can be interpreted in autonomous systems is left for future investigations.

Comparing Eqs.~\eqref{FTcoh2} and \eqref{sec:lawa} we reach the conclusion (and the third important result in this paper) that the change in entropy in the bath, within the local ME framework, is only determined by the diagonal part of the system's Hamiltonian, and only $H_D$ enters the FT \eqref{FTcoh2} and  the second law \eqref{sec:lawa}. When the ME is global, $H_D=H$,  one recovers the standard definition  $\dot  Q_{D,\alpha}=\dot  Q_{\alpha}=\tr\pg{ D_\alpha[\rho] H}$ (see Ref.~\cite{Alicki1979}).

In this respect, it is interesting to consider the case of {\it absence of coherence at $t=0$ and $t$}.
This situation occurs when the ME (\ref{me:eq}) is global, and the initial state $\rhoz$ has no coherence (i.e. the system is classical at any time), or when one performs measurements of the system state at $t=0$ and $t$, as in the two-point measurement scheme.
In this case, it makes sense to introduce the concept of classical {\it trajectories} starting from the state $\ket{j_0}$ at $t=0$ and ending in the state $\ket{j}$.
From Eq.~\eqref{FTcoh} or \eqref{FTcoh2} one then recovers immediately the classical FT \eqref{FTcl} with $\SB+\SS=-\sum_\alpha \beta_\alpha Q_\alpha+\log \rhof_{jj}-\log \rhoz_{j_0 j_0}$. 

Our results also hold when the system's Hamiltonian and/or the dissipation rates $\gamma_{\alpha,\gamma}$, entering Eq.~(\ref{dsa}), are time-dependent.
In the special case of a single bath at inverse temperature $\beta$
 and choosing the final and initial  states to be the Gibbs states $\rhof=\exp(-\beta H(t))/Z_{t}$, and $\rhoz=\exp(-\beta H(0))/Z_0$,
Eq.~\eqref{FTcoh} becomes
\begin{eqnarray}
&&\tr [ \sum_{j_0} \rhoz_{j_0j_0} \Psiu(t\,  |\, \pi_{j_0}) \E^{\beta H_{j_0j_0}(0)} \E^{-\beta H(t)}]=\nonumber \\
&&=Z_{t}/Z_0=\E^{-\beta \Delta F}\label{FTJE},
\end{eqnarray} 
where  $\Delta F$ is the difference in equilibrium free energy between the final and the initial thermal states. Furthermore, by assuming the LDBC the last equation reduces  to the JE. Indeed, following the same procedure that leads to Eq.~\eqref{FTcoh2} we find 
\begin{eqnarray}
&&\sum_{j,j_0}\int \D Q \, \E^{\beta (Q_D -H_{j j}(t)+H_{j_0j_0}(0))} \Phi_{j}(Q_D,t| \pi_{j_0}) p_{j_0}^{(0)}\nonumber\\
&&=\int \D W_D\, P(W_D)  \E^{-\beta W_D} =\E^{-\beta \Delta F} \label{FTJE2},
\end{eqnarray} 
where we have set $\SB=-\beta Q_D$, and have defined work as $ W_D\equiv\Delta H_D-Q_D$. Yet, it is worth noting that requiring the local detailed balance for a time-dependent Hamiltonian as we do in Eq.~\eqref{FTJE2}, is equivalent to requiring that the jump rates adjust instantaneously to the value of the energy level, which in turn holds only for slow driving.

\subsection{ Numerical examples}
While results \eqref{FTcoh} and \eqref{FTcoh2} are exact, we resort to numerical simulations to exemplify them and gain physical insight into their implications.
Specifically, we consider  a system of two spin-1/2 particles, characterised by Pauli operators $\sigma_{x,y,z}$, with Hamiltonian 
$H=-J \sigma_{x,a}\sigma_{x,b}-h( \sigma_{z,a}+\sigma_{z,b})$.
We consider both the cases where each spin is connected to an equilibrium reservoir at temperatures $T_a$ and $T_b$, respectively and where the two baths are non thermal, with the dissipation rates not satisfying the LDBC. The jump operators ``flip'' the individual spins: $L_\lambda=\sigma_{-,a}\otimes \II_b$ or $L_\lambda=\II_a \otimes \sigma_{-,b}$. When $J\neq 0$ the quantum ME (\ref{me:eq}) is local, see Appendix~\ref{num:app} for further details on the numerical simulations. 

We numerically evolve the system's state using the QMC algorithm for the unraveling of the ME~\cite{Molmer:93} provided by Qutip~\cite{Qutip12,Qutip13}.
The QMC algorithm evolves the system's state from $\ket{\psi(0)}$ to $\ket{\psi(t)}$ with an alternation of continuous dynamics and stochastic jumps \cite{Molmer:93}.
Along each trajectory, we monitor the individual jumps, driven by one of the two baths, and collect the statistics for the generalized bath's entropy:
\begin{equation}
\cS=\sum_{l=1}^{n_a}\Delta s_{a,j_{l+1},j_l}(t_l)+\sum_{m=1}^{n_b}\Delta s_{b,j_{m+1},j_m}(t_m),
\label{eq:SBsym}
\end{equation} 
where $t_l$ is the time at which the $l$-th jump (out of $n_a$ total jumps)  induced by the bath $a$ occurs, and analogously for bath $b$.
When the LDBC holds, the previous equation has the simple interpretation $\cS=\SB=-(\beta_a Q_{a,D}+\beta_b Q_{b,D})$.

\begin{figure*}[t]
\center
\includegraphics[width=0.18\textwidth]{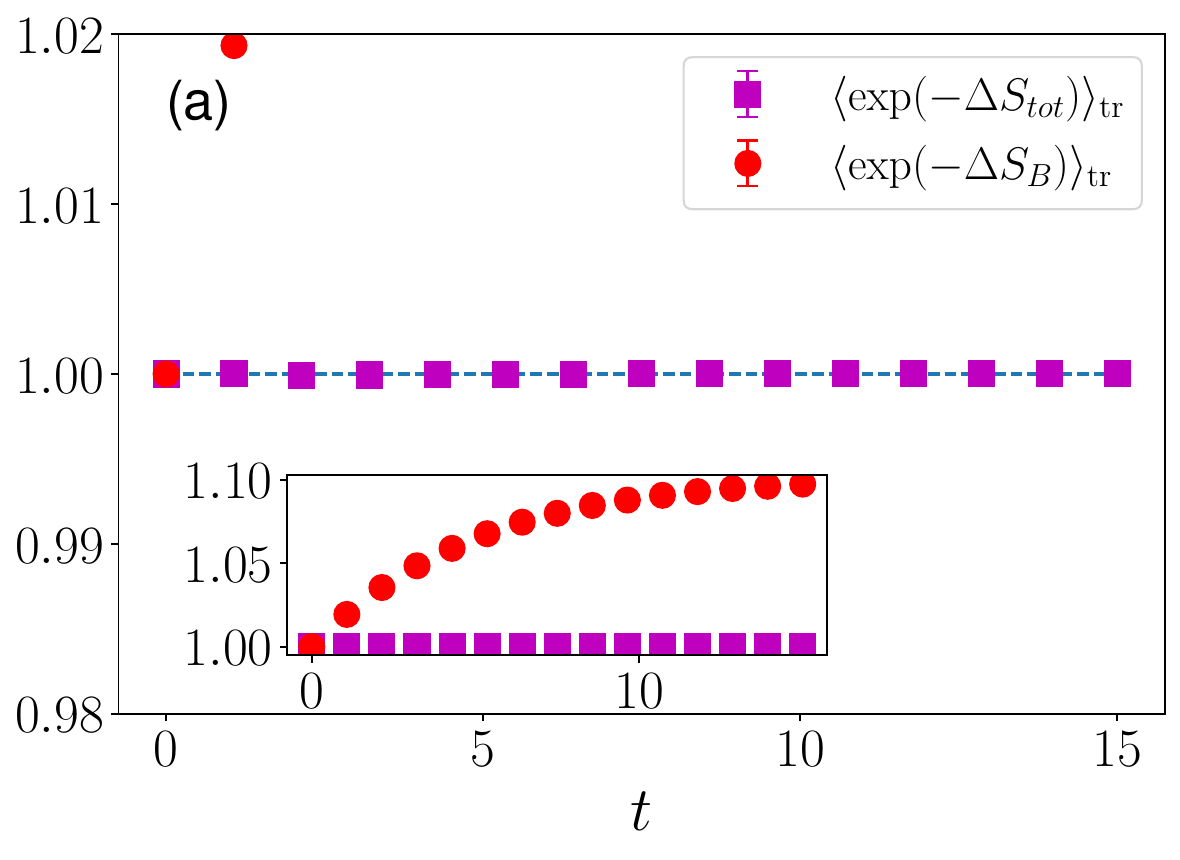}
\includegraphics[width=0.18\textwidth]{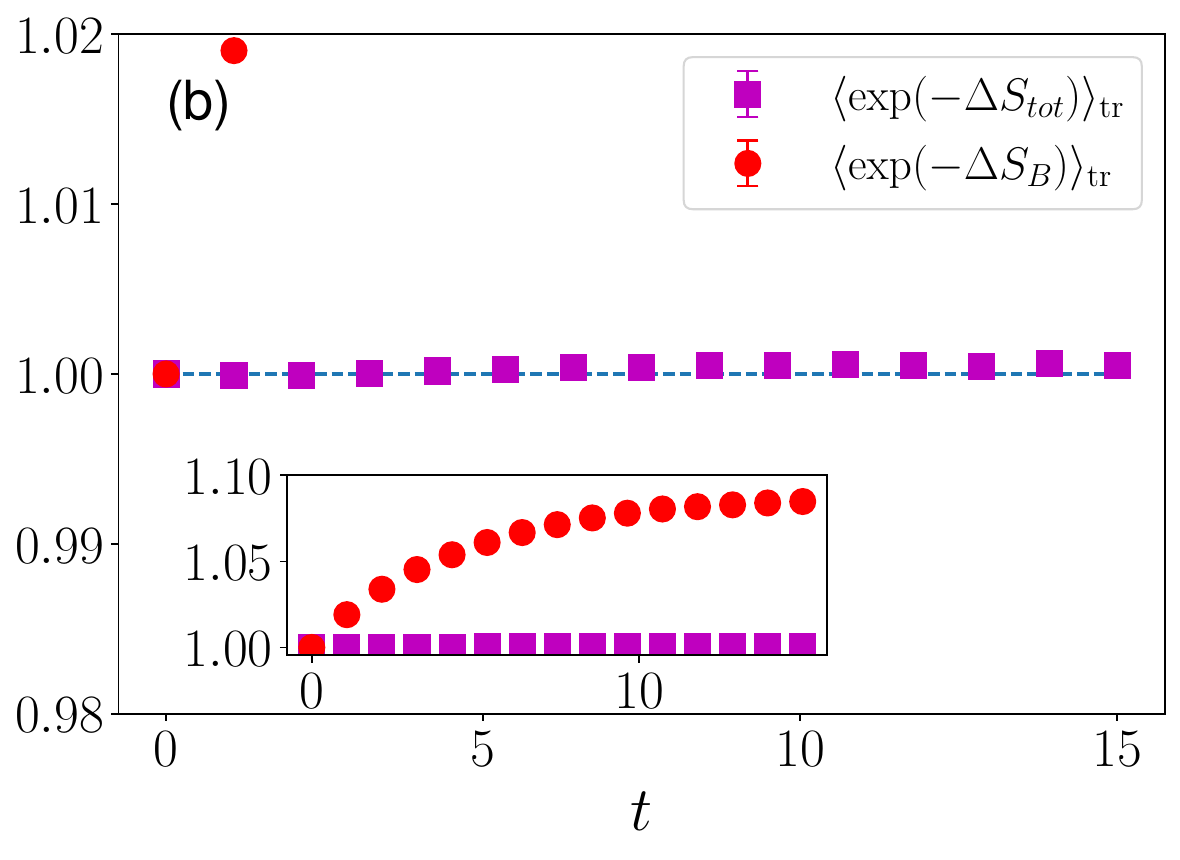}
\includegraphics[width=0.18\textwidth]{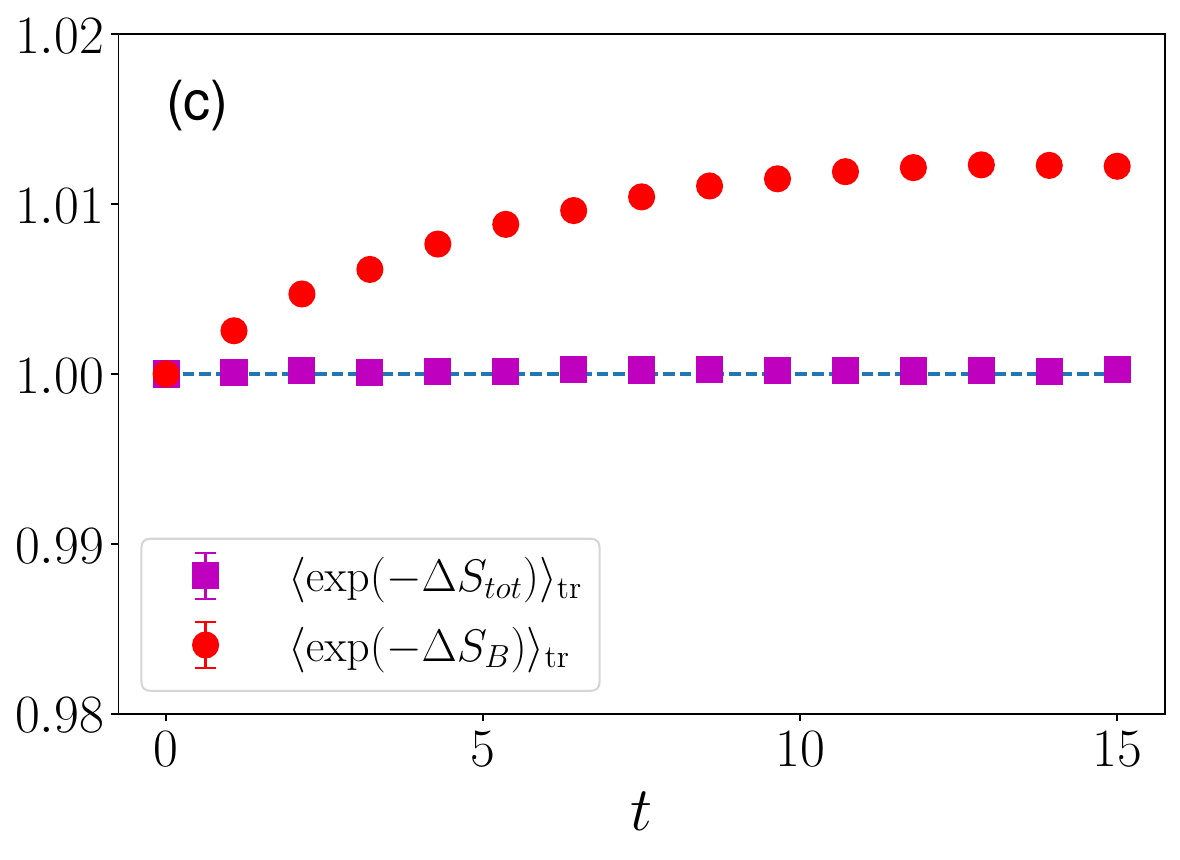}
\includegraphics[width=0.18\textwidth]{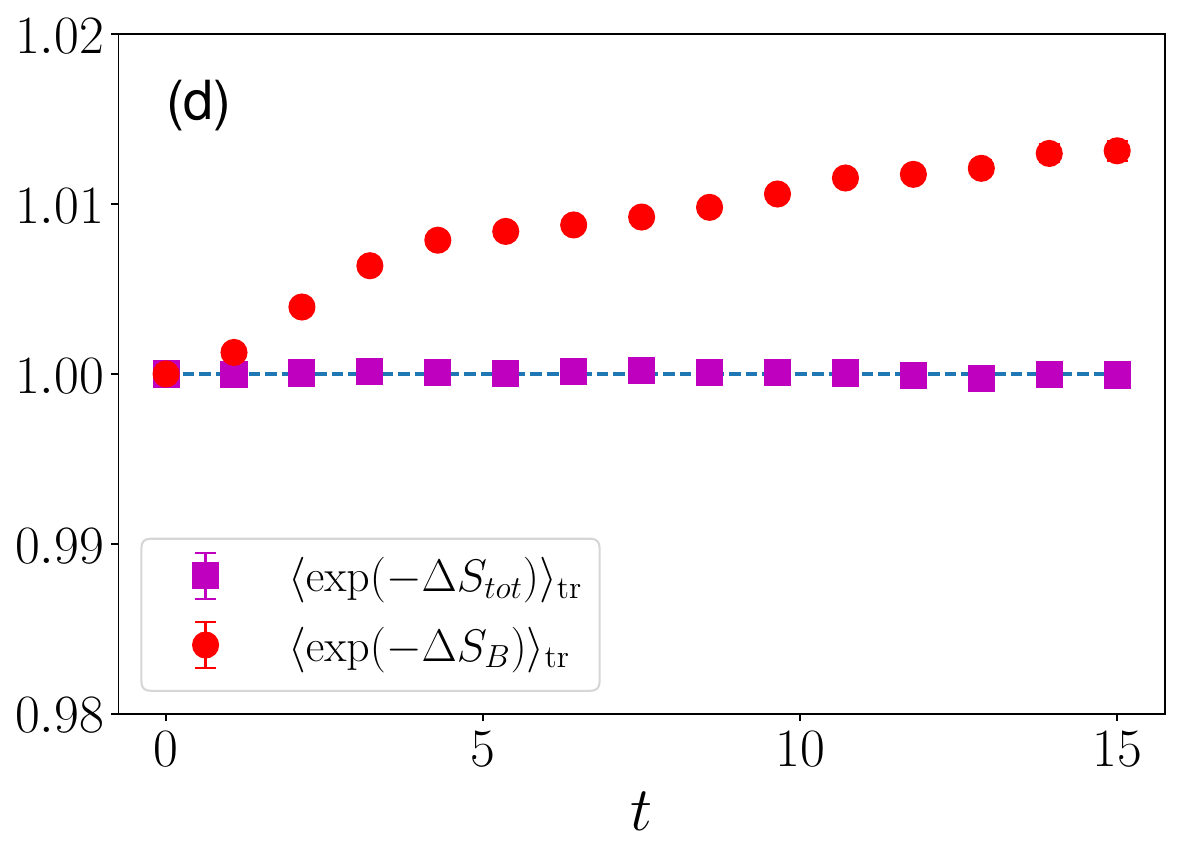}
\includegraphics[width=0.18\textwidth]{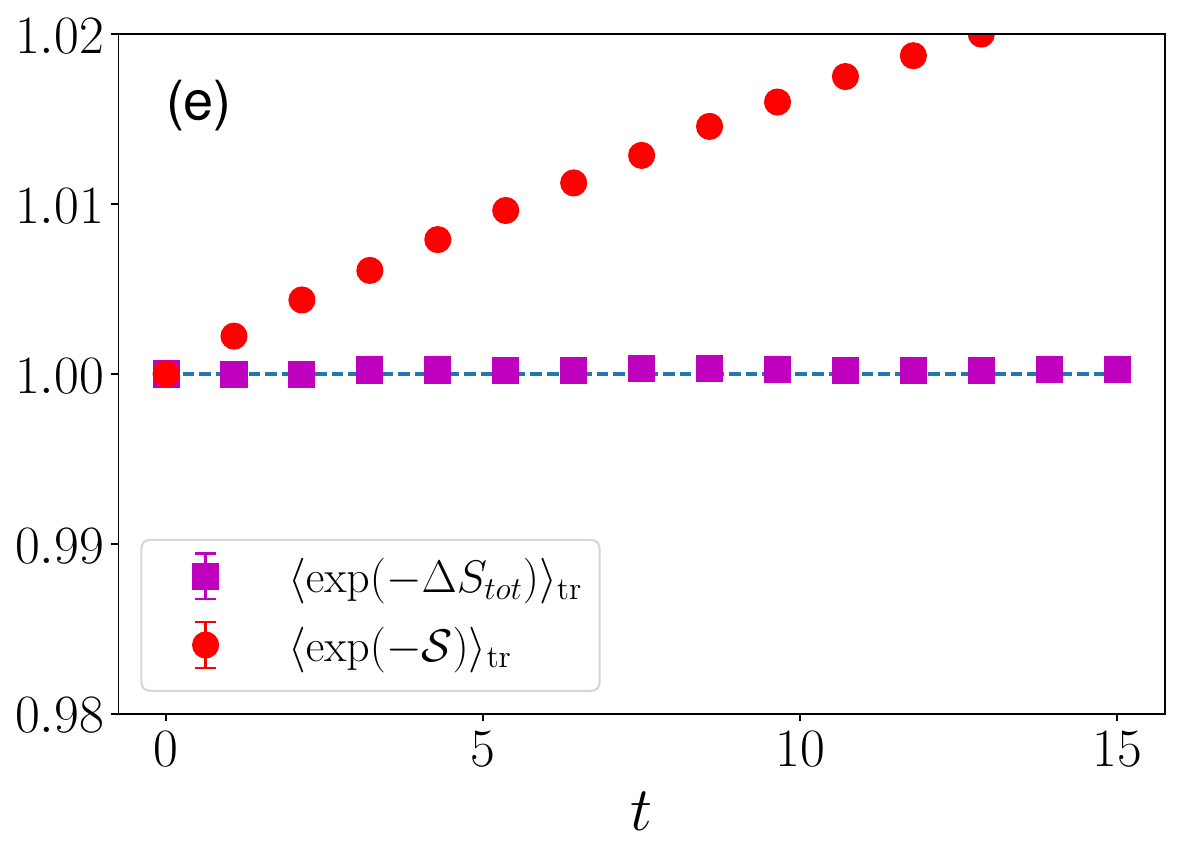}
\caption{QMC simulations of the two-spin system: $\average{\exp\p{-\Delta S_{\rm tot}}}_{\mathrm{tr}}$ and $\average{\exp\p{-\SB}}_{\mathrm{tr}}$ as functions of $t$, with with $\Delta S_{\rm tot}=\cS+\SS$. Averages over $10^6$ trajectories, $g=0.1$, $T_a=1$. (a): diagonal $\rhoz$, $J=0$, $h=0.2$,  $T_b=1$ (classical case at equilibrium). (b):  diagonal $\rhoz$, $J=0.1$, $h=0.2$,  $T_b=1.2$. (c):  nondiagonal $\rhoz$, $J=0.1$, $h=0.2$,  $T_b=1.2$. (d): nondiagonal $\rhoz$, $J=0.2$,  $T_b=1.2$, and time dependent field $h(t)=0.4 t/\tf$, with $\tf=15$. In panels (a)-(d) the LDBC is used, and thus $\cS=\Delta S_B$. (e) Non-thermal jump rates $\gamma_{\alpha,\lambda}$ that do not satisfy the LDBC, with diagonal $\rhoz$, and $J=0.1$, $h=0.2$. See Appendix~\ref{num:app}  for further details on the numerical simulations. }
\label{figmc1}
\end{figure*}

We evaluate the quantum FT in the form of Eq.~\eqref{FTcoh}  where now the average is taken over the trajectories generated by  QMC.
Within QMC, one can evaluate the solution $\rho(t)$ to the ME  (\ref{me:eq}) as 
$\rho(t)=\average{\ket{\psi(t)}\bra{\psi(t)}}_{\mathrm{ tr}}$,
where $\average{\cdots }_{\mathrm{ tr}}$ is the average over the QMC trajectories. 
Analogously, from the definition of $\Psi(\xi,t)$ in Eq.~\eqref{psi:def},  and from the sampling of the generalized entropy (\ref{eq:SBsym}) along the trajectories,  we can evaluate $\Psiu(t\,  |\, \pi_{k_0})$,  for any chosen initial basis $\{\ket {k_0}\}$, as discussed in Appendix~\ref{num:app}.

The system's entropy change $\Delta S_{S}$ can be also evaluated along the QMC trajectories as detailed in Appendix~\ref{num:app}, and we can thus evaluate $\average{\exp\p{-\Delta S_{\rm tot}}}_{\mathrm{tr}}$, with $\Delta S_{\rm tot}=\cS+\Delta S_s$.

We consider both the case where the baths are thermal, with rates $\gamma_{\alpha,\lambda}$ obeying the  LDBC, and the case of rates not satisfying the LDBC. 
Our numerical results in Fig.~\ref{figmc1} confirm the quantum FT~\eqref{FTcoh}
for initially diagonal and nondiagonal states in contact with thermal baths at the same temperatures ($T_a=T_b$) or different ones. Moreover, in Fig.~\ref{figmc1}-(d) we consider the case of a time-dependent Hamiltonian, where the external field is changed according to $h(t)=h_0+(h_1-h_0) t/\tf$.
In Fig.~\ref{figmc1}-(e) we show the results for nonthermal baths violating the LDBC.
In the same figure, we also show that, for all the considered cases,  $\average{\exp(-\SB)}\ge 1$, thus demonstrating the relevance of the system entropy variation for the FT to hold.

\section{FT and local detailed balance for unitary dynamics}
\label{uni:sec}

The aim of this section is twofold.
First we want to generalize the results (\ref{FTcoh})-(\ref{FTcoh2}) of the previous section to the case of unitary dynamics, in a setup made of  the system of interest interacting with a set of baths.
We will then rederive Eq.~\eqref{eqPsidual} governing the time evolution of the operator $\Psiu$ under the same assumptions that lead the standard ME \eqref{me:eq}.
This approach will provide the reader with a more physical intuition of the current $\cS$, that we introduced in the previous section, and that enters the FT in the form of Eqs.~\eqref{FTcoh}-\eqref{FTcoh2}.

We consider a total system made of the system of interest  interacting through a Hamiltonian $V$  with a set of
baths with Hamiltonian $H_{B_\alpha}$: the total  Hamiltonian $\Ht$ thus reads
\begin{eqnarray}
\Ht&=& \Hb + H + V, \label{Htot:def}\\
H_B&=&\sum_\alpha H_{B_\alpha},\\
V&=&\sum_{\al} V_{\al},
\label{V:def}
\end{eqnarray} 
where $ V_{\al}$ is the interaction Hamiltonian with the $\alpha$-th bath. The strength of the bath-system is taken to be arbitrary in the following discussion, the standard weak coupling limit being considered only later in this section.

We assume that initially the system and the baths are in a product state 
\begin{eqnarray}
\rt(0)&=&\rho(0)\otimes \rhob(0)\\
\rhob(0)&=& \rho_{B_1}(0)\otimes \rho_{B_2}(0)\otimes \dots \otimes \rho_{B_{N_b}}(0)
\end{eqnarray} 

Furthermore, we introduce the operators
\begin{eqnarray}
\Sba&=&-\log\rho_{B_\alpha}(0), \label{SBa:def}\\
\Sb&=&\sum_{\alpha=1}^{N_b} \Sba,
\label{SB:def}
\end{eqnarray} 
akin to an entropy operator for the single bath and for all the baths, respectively.
The case where some of eigenvalues of the $\rho_{B_\alpha}(0)$ are zero does not pose an issue on the discussion below, as we will be interested in the exponential of $\Sba$.
We do not assume, \emph{a priori}, that the initial states of the baths $\rho_{B_\alpha}(0)$ are thermal, although we will consider this specific case at a later stage.
We also assume that the baths' Hamiltonians, and the interaction Hamiltonian $V$ are time independent while the system Hamiltonian can be time dependent as discussed later.

Having introduced the total system, we start the discussion by deriving a rather obvious result, which will however set the stage for later derivations.
We introduce the time reversal operator for the total system $\theta$, such that $\comm \theta \Ht=0$ and $-\ii \Ht \theta \bullet=\theta \ii \Ht \bullet$, the latter equality expressing the antilinearity of the operator $\theta$ \cite{sakurai_napolitano_2017}.

We assume that the initial state for the total system is  a pure state $\ket {\Nu_0}$, (eigenstate of some observable). Here and in the following $\ket{\Nu_x}$ indicates states of the total system and $\Pi_x=\ket{\Nu_x} \bra{\Nu_x}$  the corresponding  projectors, while  $\ket{j_x}$ and $\pi_x=\ket{j_x}\bra{j_x}$ are states and projectors of the system of interest only as in the previous section.

The probability to observe the state $\ket {\Nu_f}$  after the total system evolves until time $t$ is 
\begin{equation}
P(  \Nu_f, t | \Nu_0, t=0)=\tr \pg{ \Pif \Ut \Pi_0 \Ut^\dagger \Pif}, 
\end{equation} 
where the unitary operator $U_t$ reads 
\begin{equation}
\Ut=\mathcal{T} \exp\pq{-\ii \int_0^t \D t' \p{\Hb+H(t')+V}}.
\label{Ut:def}
\end{equation} 
and $\mathcal{T}$ is the time-ordering operator.
We also introduce the unitary operator when the global system evolves under the time reversed Hamiltonian $\Hb+H(t-t')+V$.
\begin{equation}
\Utt=\mathcal{T} \exp\pq{-\ii \int_0^t \D t' \p{\Hb+H(t-t')+V}}.
\end{equation} 
We now assume we can prepare the total system in the state $\ket {\tilde \Nu_f}=\theta \ket { \Nu_f}$ and evaluate the probability to find the system in the state 
 $\ket {\tilde \Nu_0}=\theta \ket {\Nu_0}$ after evolving under the time reversed dynamics for a  time $t$
\begin{eqnarray}
&&P(\tilde \Nu_0, t| \tilde \Nu_f, t=0)=\tr \pg{ \tilde \Pi_0 \Utt \tilde \Pi_f \Utt^\dagger \tilde \Pi_0}\nonumber \\
&&=\tr \pg{ \Pi_0 \Ut^\dagger \Pi_f \Ut \Pi_0}= P(\Nu_f, t | \Nu_0, t=0),
\label{Psrev:tot}
\end{eqnarray} 
where we have used the antilinearity of $\theta$.
This result is not surprising and expresses the conservation of the probability under time reversal and for unitary dynamics.

We now imagine that only the system is accessible to our measures, and assume the system and the baths are initially (only initially) in a product state. Furthermore we want to do thermodynamics, and introduce a temperature (or a set of temperatures), thus we assume the bath(s) to be in the thermal state, $ \rhob(0)$.
We are then interested in the probability
\begin{eqnarray}
&&P  (j_f, t  | j_0)=\tr  \pg{ \pif \otimes \II_B \Ut \pi_0\otimes \rhob(0) \Ut^\dagger \pif\otimes \II_B}\nonumber  \\
&&=\tr_s \pg{ \pif \rho_s(t)\pif},
\label{Psfor} 
\end{eqnarray} 
and where, here and in the following, we have omitted to indicate the initial time $t=0$.
We then prepare the total system in the time-reversed state  $ \trt(0)=\theta \pi_f\otimes \rhob(0) \theta^{-1}$, and ask what is the probability that the system of interest is in the state $\ket {\tilde j_0}$ after a time $t$. We remind the reader that $\theta$ is the time reversal operator for the whole system.
A straightforward calculation gives
\begin{eqnarray}
&&P  (\tilde j_0, t  | \tilde j_f)= \tr  \pg{ \tilde \pi_0 \otimes \II_B \Utt \tpif\otimes \rhobt(0) \Utt^\dagger \tilde \pi_0\otimes \II_B}\nonumber  \\ 
&&= \tr  \pg{ \pi_0 \otimes \II_B \Ut^\dagger  \pif\otimes \rhob(0) \Ut \pi_0\otimes \II_B} \label{Psrev} \\
&& \neq P  (j_f, t  | j_0). \nonumber
\end{eqnarray} 

Compared to the case discussed above in this section, we see that the lack of symmetry in the probability under time reversal is due to the fact that we are now considering states of the system alone. In other words, we are tracing over the bath, before evaluating the transition probability of the system: this is the origin of the irreversibility in the transition probability. 
The goal is now to express the transition probability \eqref{Psrev} in terms of the forward dynamics, and find a relation between the probability for the forward and backward dynamics that can lead us to a FT, as is the case for classical dynamics \cite{Maes2021}.

In the following we will use the notation $\average{A}_{j_f,j_0}$ to indicate the time evolution of an observable with the forward dynamics, with constraints on the initial and final states.
We also notice that the two-time correlation of the observables $A$ and $B$ reads 
\begin{equation}
\average{A(t) B(0)}=\tr[\Ut^\dagger A \Ut B \rt(0)]. 
\end{equation}

Using the definition of $\Sb$ in Eq.~\eqref{SB:def}
we consider the two-time correlation 
\begin{eqnarray}
&&\average{ \E^{-\Sb(t)}  \E^{\Sb(0)}}_{j_f,j_0}=\nonumber \\
&&=\tr  \pg{ \pif \otimes \rhob(0)  \Ut   \pi_0\otimes \II_B \Ut^\dagger \pif\otimes \II_B}\nonumber \\
&&=P  (\tilde j_0, t  | \tilde j_f).
\label{quantum:crook}
\end{eqnarray} 

This is a generalization  of the  Crooks' quantum fluctuation relation \cite{Crooks2008} to the case of arbitrary interaction strength between the system and the baths, and for an arbitrary number of baths. Such a generalization  relates the probability of the time reversed transition to the two time correlation function of the  exponential of the baths' entropy. For the classical case, with each  bath {\it initially} in a thermal state at temperature $\beta_\alpha$, Eq.~\eqref{quantum:crook} becomes  
\begin{equation}
P  ( j_f, t  |  j_0) \E^{- \Delta S_B} =P  (\tilde j_0, t  | \tilde j_f)
\end{equation} 
and $\Delta S_B=\sum_\alpha \beta_\alpha \Delta \Hba $ is the change of entropy in the baths, along any system trajectory connecting $ j_0$ to $j_f$.

From Eq.~\eqref{quantum:crook} we obtain immediately 
\begin{eqnarray}
&&\sum_{j_0} \average{ \E^{-\Sb(t)}  \E^{ \Sb(0)}}_{j_f,j_0}= 
\label{FTint}
\\
&&=\sum_{j_0}\tr  \pg{ \pif \otimes \rhob(0)  \Ut   \pi_0\otimes \II_B \Ut^\dagger \pif\otimes \II_B}=1.
\nonumber
\end{eqnarray} 

We then  continue along the lines of the previous section 
and consider the arbitrary basis  $\{\ket{b_0}\}$  and arbitrary initial and final states $\rhoz$ and $\rhof$, and from Eq.~\eqref{FTint} we obtain 
\begin{eqnarray}
&&\sum_{b_0} \average{\E^{\log \rhof(t)} \E^{- \Sb(t)}  \E^{ \Sb(0)} \E^{ -\log \rhoz_{b_0,b_0}}}_{b_0}\,  \rhoz_{b_0,b_0}=\nonumber \\
&&=\sum_{b_0}\tr  \pg{ \rhof \otimes \rhob(0)  \Ut   \pi_{b_0} \otimes \II_B \Ut^\dagger \pif\otimes \II_B}\nonumber\\
&&= \sum_{b_0} P  (\tilde b_0, t  | \rhoft)
 =1.
\label{FTun}
\end{eqnarray}

Eqs.\eqref{FTint}-\eqref{FTun} represent the FT for the joint unitary evolution of the system and the baths: they generalise Eqs.~\eqref{FTcoh}-\eqref{FTcoh2} obtained starting from the master equation of the system alone.
This result for the global unitary dynamics involves the two-time correlation of the entropy in the bath, which is related to the backward dynamics as expressed by Eq.~\eqref{quantum:crook}.
Furthermore Eq.~(\ref{FTun}) involves the two-time correlation function of the system entropy too. Yet an initial measurements on the system is required, represented by the projection onto the arbitrary basis state $\ket{b_0}$.
As discussed in the previous section, while the final state $\rhof$ is arbitrary, one can of course take the specific choice $\rhof=\rho(t)=\tr_B \rt(t)$, i.e. the final state of the system along the forward dynamics. In this case, the FT involves the change of entropy in the system.

In deriving the above results we explicitly assumed that the system Hamiltonian is time dependent $H(t)$ while the bath and the interaction Hamiltonian are time independent. 
Thus, even when work is done on or by the system on the external agent changing the system Hamiltonian, the relevant quantities to consider in the FT are the correlation between the exponential of the baths' entropy, as appearing on the right-hand side (RHS) of Eqs.~\eqref{FTint} and \eqref{FTun}.
As such, these equations do not require the monitoring of the system energy change, or of the work done on or by the system, that might disturb the quantum dynamics of the system. Such an approach, requiring the monitoring of the environment only, was already put forward in, e.g., \cite{Crooks2008}.

\subsection{Connection to the modified ME}
\label{ME:subsec}

We now connect the fluctuation relations for the unitary dynamics, Eqs.~\eqref{quantum:crook}-\eqref{FTun} with the modified master equation \eqref{eqPsidual} introduced and discussed in the previous section,  (see also Eq.~\eqref{PsiND:app} in Appendix~\ref{mod:app}).

For the interaction Hamiltonian between the system and the $\alpha$-th bath, introduced in Eq.~\eqref{V:def}, we choose   the general form
\begin{equation}
V_{\al}= g_{\al} (L^\dagger_\lambda  A_{\al}+ L_\lambda A^\dagger_{\al}), 
\label{V:alpha}
\end{equation} 
where  $L_\lambda$ are system operators, and $A_{\alpha,\lambda}$ are bath operators.

In the following we make a number of assumptions on the baths' properties and on the system-baths interaction, that are usually introduced in the derivation of  the standard quantum ME \eqref{me:eq} \cite{Cohen_Tannoudji}.
In particular we make the  requirement that the baths' initial states commute with their corresponding  Hamiltonians $\comm{\Hba}{\rho_{B_\alpha}(0)}=0$.
Furthermore, we choose the bath operators to be eigenoperators of the bath entropy operator 
\footnote{Given the assumption that the bath Hamiltonian commutes with the corresponding initial state this is equivalent to assume that $A_{\al}$ are eigenoperators of  $\Hba$.}
\begin{equation}
\comm{\Sba}{A_{\al}}=\Omega_{\al} A_{\al}.
\label{comm:SA}
\end{equation} 
It is always possible to express the interaction Hamiltonian \eqref{V:alpha} in terms of eigenoperators of $\Sba$. Indeed if $A_{\al}$ are not  eigenoperators of $\Sba$, by expressing them in terms of the eigenbasis of $\Sba$ one can write $V_\alpha$ in terms of new (rotated) operators $A'_{\al}$ which are eigenoperators of $\Sba$. In order to ease the notation, in the following we consider the case where the system Hamiltonian is time independent, the derivation of the corresponding results for the time dependent case follows the same lines discussed below.

By using the assumption in Eq.~\eqref{comm:SA}, one can show the following equality to hold
\begin{eqnarray}
\E^{-\frac{S_B} 2}\E^{-\ii t \Ht}=\E^{-\ii t \bar \Ht} \E^{-\frac{S_B} 2}
\label{comm:tilde}
\end{eqnarray} 
with 
\begin{eqnarray}
\bar \Ht =H_B+ H + \bar V
\end{eqnarray} 
and where 
\begin{eqnarray}
\bar V&=&\sum_{\al} \bar V_{\al}
\label{Vtilde}
\\
&=&\sum_{\al}g_{\al} \p{ \E^{-\beta_\alpha \Omega_{\al}/2} L^\dagger_\lambda A_{\al} + \E^{\beta_\alpha \Omega_{\al}/2} L_\lambda A^\dagger_{\al}   }
\nonumber
\end{eqnarray} 
is a modified, non-Hermitian interaction Hamiltonian, see Appendix~\ref{unitary:ME:app}.

Let $\bar U_t$ be the nonunitary operator obtained from  Eq.~\eqref{Ut:def} with the substitution $V\to \bar V$.
Then comparing \eqref{comm:tilde}  with \eqref{quantum:crook}, we finally find 
\begin{eqnarray}
&&P  (\tilde j_0, t  | \tilde j_f, t=0)=\average{ \E^{-\Sb(t)}  \E^{ \Sb(0)}}_{j_f,j_0}\nonumber \\
&& =\tr  \pg{ \pi_f \otimes \II_B \bar U_t \pi_0\otimes \rhob(0) \bar U_t^\dagger \pi_f\otimes \II_B},
\label{Psrev:tilde}
\end{eqnarray} 
resembling Eq.~\eqref{Psrev:tot} that expresses the conservation of probability under time reversal for the total system.
However, in \eqref{Psrev:tilde} the forward time evolution occurs with  $\bar U_t$.

Let us now consider the evolution of the system operator 
\begin{eqnarray} 
\Psi(t+\tau)\equiv \tr_B  \pg{\bar U_\tau \Psi(t)\otimes \rhob(0) \bar U_\tau^\dagger },
\label{Psrev:tilde1}
\end{eqnarray} 
given an arbitrary (possibly mixed) initial state  of the system $\Psi(t)$.
The projection  of such an operator on the state $j_f$, is the two-time correlation function  after a  time $t+\tau$, as  introduced in Eqs.~\eqref{quantum:crook} and \eqref{Psrev:tilde}, given the system state at time $t$ $\Psi(t)$ 
\begin{equation}
\average{ \E^{- \Sb(t+\tau)}  \E^{ \Sb(t)}}_{j_f,\Psi(t)}=\bra{j_f} \Psi(t+\tau) \ket{j_f}.
\label{rhost:def}
\end{equation} 
We make the Born-Markov approximation valid in the  weak coupling limit, and assume that the global system is in a product state at any time  $\rt=\Psi(t)\otimes \rhob$, and that the baths' state is time independent $\rhob=\rhob(0)$ \cite{Cohen_Tannoudji, Breuer02}. Thus in Eq.~\eqref{Psrev:tilde1}
we take the expansion 
 up to the second order in $\tau$: some extra care is needed as $\bar V$ (see Eq.~\eqref{Vtilde}) and thus $\bar \Ht$, is not Hermitian. We thus obtain 
\begin{eqnarray}
 \Psi(t+\tau)&=& \tr_B \pg{\bar U_\tau \Psi(t) \otimes \rhob \bar U_\tau^\dagger}
 \nonumber \\
& \simeq &\tr_B \left \{ \p{\II -\ii \tau \bar \Ht -\frac{\tau^2} 2 \bar V^2}  \right .
\label{rhost:exp}
\\
 &&\left . \Psi(t) \otimes \rhob \p{\II +\ii \tau \bar \Ht ^\dagger -\frac{\tau^2} 2( \bar V^\dagger)^2} \right\}
 \nonumber 
\end{eqnarray} 
A lengthy but straightforward calculation gives
\begin{widetext}
\begin{eqnarray}
 \Psi(t+\tau)&\simeq& \Psi(t) -\ii \tau [\Hs,\Psi(t)] + \tau^2 \sum_{\al} g_{\al}^2 \p{   \E^{\beta_\alpha \Omega_{\al}}  L_\lambda \Psi(t) L_\lambda^\dagger\average{ A_{\al}  A_{\al}^\dagger}_B + \E^{-\beta_\alpha \Omega_{\al}} L_\lambda^\dagger \Psi(t) L_\lambda\average{ A_{\al}^\dagger  A_{\al}}_B  }\nonumber \\
&& -\frac{\tau^2} {2}  \sum_{\al} g_{\al}^2 \p{  \pg{  L_\lambda^\dagger  L_\lambda, \Psi(t) } \average{ A_{\al}  A_{\al}^\dagger}_B +  \pg{  L_\lambda  L_\lambda^\dagger, \Psi(t) } \average{ A_{\al}^\dagger  A_{\al}}_B}.
\label{rhost}
\end{eqnarray} 
\end{widetext}

We now introduce the rates 
\begin{eqnarray}
\gamma_\alpha(j \to j')&=&\tau g^2_{\alpha,\lambda} \average{A_{\alpha, \lambda} A^\dagger_{\alpha, \lambda}}_{B},
\label{gammajj1}\\
\gamma_\alpha(j' \to j)&=&\tau g^2_{\alpha,\lambda} \average{A^\dagger_{\alpha, \lambda} A_{\alpha, \lambda}}_{B}. 
\label{gammaj1j}
\end{eqnarray}

We see that, by virtue of Eq.~\eqref{comm:SA}, $A_{\alpha, \lambda} A^\dagger_{\alpha, \lambda}$ and $A^\dagger_{\alpha, \lambda} A_{\alpha, \lambda}$ are projectors onto the eigenstates of $\Sba$ \eqref{SBa:def}. Furthermore, by taking $A_{\alpha, \lambda}=\ket{N'_\alpha} \bra{N_\alpha}$, we have that the quantity $\Omega_{\al}$ introduced in Eq.\eqref{comm:SA} becomes an entropy gap $\Omega_{\al}=-(s_{N'_\alpha}- s_{N_\alpha})=\log r_{N'_\alpha}- \log r_{N_\alpha}$, where $r_{N_\alpha}$ are the eigenvalues of the bath state $\rho_{B_\alpha}$.
Thus, we find that the jump rates for the system dynamics appearing both in the system ME~\eqref{me:eq} and in the modified ME \eqref{eqPsidual}  obey the following relation 
\begin{equation}
\frac{\gamma_\alpha(j \to j')}{\gamma_\alpha(j' \to j)}=\frac{\average{A_{\alpha, \lambda} A^\dagger_{\alpha, \lambda}}_{B}}{\average{A^\dagger_{\alpha, \lambda} A_{\alpha, \lambda}}_{B}}=  \E^{\Omega_{\al}},
\label{dbB:eq}
\end{equation} 
We remind the reader that, given our choice of the interaction Hamiltonian \eqref{V:alpha} and the jump operators $L_\lambda$ and $A_{\al}$, a jump $\lambda$  in the system $j\to j'$ corresponds to a jump in the bath  $N'_\alpha \to N_\alpha$.
Eq.~\eqref{dbB:eq} determines the asymmetric part of the system rates,
and involves the entropy change in the bath for a transition $(j\to j', N'_\alpha \to N_\alpha)$, or equivalently the ratio of the population between the two states $N_\alpha$ and $N'_\alpha$ in the initial bath state.
Neither the entropy nor the energy of the system appear in this relation.

Let $\epsilon_j=H_{jj}$ be the diagonal elements of the system Hamiltonian in the basis $\pg{\ket j}$. Thus the quantities $\epsilon_j$ are eigenvalues of $H$ only if the chosen basis is an eigenbasis of $H$.
Eq.~\eqref{dbB:eq} becomes the standard {\it local detailed balance condition} involving the {\it system} energy gaps introduced in Sec.~\ref{fluct:theo:ME}
$\gamma_\alpha(j \to j')/\gamma_\alpha(j' \to j)=\exp(-\beta_\alpha \omega_\lambda)$ only 
under two additional conditions: (i) the baths are initially in the thermal state, thus $\Omega_{\al} =-\beta_\alpha( E_{N'_\alpha}- E_{N_\alpha})$, where $E_{N_\alpha}$ are the eigenvalues of $\Hba$, and (ii) the energy gaps in the baths and in the system are {\it resonant}, i.e., $ E_{N'_\alpha}- E_{N_\alpha}=\omega_\lambda=H_{j'j'}-H_{jj}$.
We call such a condition local, because we envisage the typical physical setup where a bath is locally connected to a subpart of the system, possibly a single particle, through some mechanism as embodied by the interaction Hamiltonian \eqref{V:alpha}.
Condition (i) reminds us that the temperature (or the energy scale $k_B T_\alpha$) is a physical quantity associated with a bath at thermal equilibrium. Condition (ii) leads to the {\it global} detailed balance condition if, within the weak coupling assumptions, no energy is stored in the coupling mechanism.  We see immediately that this is the case  when $\pg{\ket j}$ is an eigenbasis for $H$:  condition (ii) is then equivalent to assume that, given a change of energy in the system, there is a corresponding change of energy in the bath, with no energy released by  or stored in the interaction mechanism represented by the Hamiltonian $V$, see also Appendix \ref{curr:app}. 

Having settled the connection between the system dynamic evolution and the underlying thermodynamic processes, as represented by Eq.~\eqref{dbB:eq}, we continue our analysis of Eq.~\eqref{rhost}.  Using the definition of the rates \eqref{gammajj1}-\eqref{gammaj1j}, the detailed balance condition \eqref{dbB:eq}, and taking  the limit $\tau\to 0$, we obtain , 
\begin{equation}
\partial_t \Psi(t) =-\ii [H,\Psi(t)]+ \sum_\alpha D_\alpha^*[\Psi(t)],
\label{gen:ME:un}
\end{equation}
i.e. the generalized ME Eq.~(\ref{eqPsidual})  (see (A.9) and Appendix~\ref{unitary:ME:app} for the details).
Had we taken the standard unitary operator $U_t$ in Eq.~(\ref{rhost:exp}) (i.e. $\bar V$ replaced by $V$ on the RHS of Eq.~(\ref{rhost:def}) ) we would have obtained the standard master equation \eqref{me:eq} in the same limit.

Thus comparing this last result with  Eqs.~\eqref{Psrev:tilde1}, \eqref{rhost:def}, we conclude that the operator $\Psiu(t)$ appearing in   Eq.~\eqref{eqPsidual} describes the joint evolution of the system state, and of the two time correlation of the baths' entropy.  For a single transition $(j\to j', N'_\alpha\to N_\alpha)$, the change in the bath entropy is given by $\Omega_{\al}$. 
We recall the definition of $\Psiu(t)=\Psi(\xi=1,t)$, Eq.~\eqref{psi:def}, and thus the quantity $\cS$ introduced in Sec.~\ref{fluct:theo:ME} is the total entropy change of the baths alone along a trajectory of the total system, the contribution of a single jump being set by the condition Eq.~(\ref{dbB:eq}).
If the baths are initially thermal, and conserve their state for the entire duration of the system dynamics, $\Omega_{\al}=-\beta_\alpha (E_{N'_\alpha}-E_{N_\alpha})$, and thus Eq.~(\ref{dbB:eq}) implies that the quantity $\cS$ is the change in the baths' energy multiplied by the corresponding inverse temperature: $\cS=\sum_\alpha \beta_\alpha \Delta H_\alpha$. 

Even if the system Hamiltonian is time dependent, i.e. work is done on or by the system, making the standard assumption that the baths are at equilibrium implies that the FT \eqref{FTcoh}, entails only the change of energy in the baths, which is of course affected by the work done on the system.

\section { Conclusions} 

We have proved a quantum fluctuation theorem valid for a quantum system evolving with a dissipative process and with a time dependent Hamiltonian. Our theorem goes beyond the two-point measurement scheme by preserving quantum correlations created by the nonequilibrium dynamics. Specifically, our results are also valid for local master equations and their physical consistency is restored when one discerns between the entropic and energetic balance between the baths, the system and the interaction mechanism. Our theorem provides a convenient tool for the numerical and experimental exploration of irreversible quantum coherent thermodynamics. 

We have also shown that the derivation of the  FT based on the ME can be extended to the case of unitary dynamics, when the explicit interaction between the systems and the bath is taken into account. This approach has the advantage of highlighting the quantity that needs to be {\it measured} in a possible experimental validation of the theorem: the two-time correlation function of the exponential of the baths' entropy.
In the case of weak coupling between the system and the baths, under the standard assumptions that lead to the GKSL ME, such a correlation function becomes a function of the jump rates appearing in the ME.

Furthermore, starting from the joint unitary dynamics of the  system and the environment has the advantage of immediately setting the connection between the system dynamics and the underlying thermodynamic processes.  The ratio between forward and backward jump rates originally depends on the properties of the baths alone:  whether this relation leads to the global, local or even to no detailed balance condition, 
 depends on further assumptions on the baths' and system's properties.
Yet the FT does not require any detailed balance condition for the jump rates appearing in the ME. 

The experimental validation of our theorem requires the monitoring of the environments and the observation of the corresponding quantum jumps. This can be achieved, for instance, in circuit and cavity-QED \cite{Haroche2020}, trapped-ion setups \cite{Schindler2013} or noisy intermediate-scale quantum computers \cite{Melo2022}.

\begin{acknowledgments}
 G.D.C. acknowledges support by the UK EPSRC EP/S02994X/1. The numerical results presented in this work were obtained at the Centre for Scientific Computing, Aarhus. All data created during this research is openly available from QUB-Pure at \url{https://doi.org/10.17034/b3e856e6-87dc-4849-8c0b-5785856cb618}

A.I. gratefully acknowledges the financial
support of The Faculty of Science and Technology at Aarhus
University through a Sabbatical scholarship and the hospitality of the Quantum Technology group, the Centre for
Theoretical Atomic, Molecular and Optical Physics and the
School of Mathematics and Physics, during his stay at
Queen’s University Belfast.

This research was supported in part by the International Centre for Theoretical Sciences (ICTS) through the online program ``Classical and Quantum Transport Processes : Current State and Future Directions"        (code: ICTS/ctqp2022/1).
\end{acknowledgments}

\appendix
\section{The modified quantum master equation}
\label{mod:app}
Here we discuss the derivation of Eq.~(\ref{rhohD}) in the main text.
Such a derivation follows the corresponding derivation of the modified ME  for classical stochastic systems discussed in Ref.~\cite{Imparato7a}.
The time evolution of $\rho(t)$ is described by the GKSL ME (\ref{me:eq}).
We notice that the diagonal terms of the density matrix $\rho(t)$ express the time evolution of the populations, i.e., the probability $p_j(t)=\rho_{jj}(t)$ of observing the system in the state $\ket j$ at time $t$.
We are interested in the joint probability distribution $\Phi_j(\cS,t)$ of finding the system in the state $j$ with a total generalised entropy $\cS$ which has flowed into the reservoirs at time $t$ as a consequence of the jumps between states in the system.
In the main text we have already introduced the  modified density matrix $\bar \rho(\cS,t)$, such that its diagonal elements describe the desired joint probability $\bar \rho_{jj}=\Phi_j(\cS,t)$.
We first observe that the dissipators in the ME, as defined in Eq. \eqref{diss}, only couple diagonal elements of the  density matrix with diagonal elements and nondiagonal elements with nondiagonal elements. The coupling between the diagonal and nondiagonal elements of $\rho$ occurs through the coherent part of the dynamics, expressed by the commutator in Eq.~\eqref{me:eq}.
Thus we can write 
\begin{equation}
D_\alpha[\rho]=D_{D,\alpha}[\rho_D]+D_{ND,\alpha}[\rho_{ND}],
\end{equation} 
where 
$D_D[\rho_D]$ and $D_{ND}[\rho_{ND}]$ contain only diagonal and nondiagonal entries in their matrix representation, respectively, and $D_D[\rho_{ND}]=D_{ND}[\rho_D]=0$.
This becomes evident when one rewrites Eq.~(\ref{me:eq}) as 
\begin{eqnarray}
\dt{ \rho(t)}&=&-\ii \pq{H(t),\rho(t)}\nonumber \nonumber \\
&&+\sum_{\alpha,\lambda} \gamma_{\lambda,\alpha} \Bigl( \ket{j'}\bra{j} \rho\ket{j}\bra{j'}  -\frac{1}{2} \p{\pi_j\rho+\rho\pi_j}\Bigr) \nonumber \\
\label{me:eq:app}
\end{eqnarray} 
where we remind the reader that $\lambda=\lambda( j\to j')$ denotes a transition between two states $\ket j$ and $\ket{ j'}$.
Isolating the diagonal part of (\ref{me:eq:app}), one obtains
\begin{equation}
\partial_t  \rho_{j'j'}=-\ii [H, \rho]_{j'j'} + \sum_{\alpha,j}
    \left\{ \gamma_{\alpha,j'j} \rho_{jj}  -\gamma_{\alpha,jj'} \rho_{j'j'}\right\}.\label{rhoD:app}
\end{equation} 

The jump between the states $\ket j$ and $\ket{ j'}$ occurs with rate $\gamma_\alpha(j\to j')=\gamma_{\alpha,j'j}$, and  the corresponding  entropy change in the bath $\alpha$ is given by 
\begin{equation}
\Delta s_{\alpha,j'j}= -\log(\gamma_{\alpha,jj'}/\gamma_{\alpha,j'j}).
\label{dsa:app}
\end{equation} 
which corresponds to Eq.~(\ref{dsa}) in the main text. Such jumps in the ME (\ref{me:eq}) are described by $D_D[\rho_D]$ implicitly introduced in Eq.~(\ref{rhoD:app}),  thus only the coupling between the diagonal elements in the ME contributes to the change in the bath generalised entropy $\cS$. After the jump $j\to j'$ the generalised entropy changes as $\cS \to \cS +\Delta s_{\alpha,j'j}$. 
 Let us now suppose that we know the modified  density matrix $\bar \rho(\cS,t)$ at time $t$, and let us consider the time evolution of its diagonal and nondiagonal parts. For the diagonal part, at time $t+\tau $, we have
\begin{eqnarray}
&&\bar \rho_{j'j'}(\cS,t+\tau)\simeq  \bar \rho_{j'j'}(\cS,t) -\ii \tau [H,\bar \rho(\cS)]_{j'j'} \nonumber \\
&&   +\tau \sum_{\alpha,j}   \gamma_{\alpha,j'j}\bar \rho_{jj}(\cS-\Delta s_{\alpha,j'j},t)-\gamma_{\alpha,jj'}\bar \rho_{j'j'}(\cS,t).
\label{rhohD:app}
\end{eqnarray} 
For the nondiagonal part of the dynamics, that does not contribute to $\cS$, we have 
\begin{eqnarray}
\bar \rho_{lk}(\cS,t+\tau)&\simeq&\bar \rho_{lk}(\cS,t)-\ii \tau   [H,\bar \rho(\cS)]_{lk} \nonumber \\ 
&+&   \tau \sum_\alpha D_{ND}[\bar \rho_{ND}(\cS,t)]_{lk}, \, l\neq k. \label{rhohND:app}
\end{eqnarray} 
Taking the limit $\tau\to 0$ in both (\ref{rhohD:app}) and (\ref{rhohND:app}), and expanding the right-hand side of Eq.~(\ref{rhohD:app}) in Taylor series of the microscopic entropy change $\Delta s_{\alpha,j'j}$, we obtain the modified quantum ME Eqs.~(\ref{rhohD})--(\ref{rhohND}) in the main text.

We  introduce the operator $\Psi(\xi,t)$  defined as
\begin{equation}
\Psi(\xi,t)=\int\, d \cS \, \bar \rho(\cS,t) \E^{-\xi \cS},
\label{psi:def:app}
\end{equation} 
and applying this integral transform to both sides of Eqs.~\eqref{rhohD} and \eqref{rhohND} in the main text, one obtains the equations for $\Psi(\xi,t)$
\begin{eqnarray}
  &&  \partial_t \Psi_{j'j'}(\xi,t)=-\ii [H,\Psi]_{j'j'}+\nonumber \\
&&\quad \sum_{\alpha,j}
    \left\{ \gamma_{\alpha,j'j} \p{\frac{\gamma_{\alpha,jj'}}{\gamma_{\alpha,j'j}}}^\xi\Psi_{jj}(\xi,t)   -\gamma_{\alpha,jj'}\Psi_{j'j'}(\xi,t)\right\},\label{PsiD:app}\\
&&\partial_t \Psi_{lk}(\xi,t)=-\ii [H,\Psi]_{lk}+D_{ND}[\Psi_{ND}]_{lk}, \quad l\neq k. \label{PsiND:app}
\end{eqnarray}

We now take $\xi=1$ in  Eqs.~(\ref{PsiD:app})--(\ref{PsiND:app}), yielding
\begin{eqnarray}
    \partial_t \Psiu_{j'j'}(t)&=&-\ii [H,\Psiu]_{j'j'}\nonumber \\ 
  &+&\sum_{\alpha,j}
    \gamma_{\alpha,jj'} \left\{  \Psiu_{jj}(t)   -\Psiu_{j'j'}(t)\right\},\label{PsiD1:app}\\
\partial_t \Psiu_{lk}(t)&=&-\ii [H,\Psiu]_{lk}+D_{ND}[\Psiu_{ND}]_{lk}, \quad l\neq k. \nonumber \\  \label{PsiND1:app}
\end{eqnarray} 

The dual of the dissipator $D_\alpha[\cdot]$ applied on an operator $X$ reads 
\begin{equation}
D_\alpha^*[X]=\sum_{j \to j''} \gamma_{\alpha,j''j}\left( X_{j''j''} \pi_j    -\frac 1 2 \{ \pi_j,X\}\right),
\label{diss:du}
\end{equation} 
and considering the diagonal element
\begin{equation}
D_\alpha^*[X]_{j'j'}=\sum_{j''} \gamma_{\alpha,j''j'}\left( X_{j''j''}     -X_{j'j'}\right),
\label{diss:du:diag}
\end{equation} 
one finds that it corresponds to the term on the right-hand side of Eq.~(\ref{PsiD1:app}). For the nondiagonal part, one finds $D_{\alpha,ND}^*[X_{ND}]=D_{\alpha,ND}[X_{ND}]$.
All in all, one finds that $\Psiu(t)$ obeys Eq.~(\ref{eqPsidual}) in the main text.

\section{On the heat exchanged between the system and the bath(s)}
\label{curr:app}
As in the main text, we consider the total system made of the system of interest with Hamiltonian $H$ and a set of baths with Hamiltonians $\Hba$, interacting through a Hamiltonian $V$, see Eqs.~\eqref{Htot:def}--\eqref{V:def}.
Given an arbitrary  orthonormal basis $\{\ket j\}$ for the system, we split its Hamiltonian into its diagonal and nondiagonal parts
\begin{eqnarray}
H&=&\HD+\HND. \label{HS:def}
\end{eqnarray} 
The time evolution of the total density matrix is thus given by
\begin{equation}
\rt(t+\tau)=\E^{-\ii \tau \Ht} \rt(t)  \E^{\ii \tau \Ht}, 
\label{rhot}
\end{equation} 
for any $t$ and $\tau$.
To lighten the notation, here and in the following we assume that all the above Hamiltonians, in particular $V$, are time independent.  Here we do not consider the case where $H$ depends explicitly on the time, since in this section we are only interested in the energy exchange between the system and the baths, and not on the work done on the system by an external agent. However the case of a time dependent $H(t)$ only requires a straightforward modification. 
As discussed in the main text, we do not assume \emph{a priori} that the states $\ket j$ are eigenstates of the system Hamiltonian.

We recall the definition of the interacting Hamiltonians introduced in Eq.~\eqref{V:def} $V_{\al}= g_{\alpha, \lambda}(L^\dagger_{ \lambda} A_{\alpha, \lambda} + L_{ \lambda} A^\dagger_{\alpha, \lambda})$,
 with the system operators $L_{\lambda}$  and the bath operators  $A_{\alpha,\lambda}$ satisfying 
\begin{eqnarray}
\pq{\HD, L_{ \lambda}}&=&\omega_{\lambda} L_{ \lambda},\label{comm:S}\\
\pq{\Hba, A_{\al}}&=&\myO_{\al } A_{\al}\label{comm:B}.
\end{eqnarray} 
with 
\begin{eqnarray}
	\omega_{ \lambda}&=&\omega_{ j'j}=\bra{j'} H \ket {j'}-\bra{j} H\ket {j}=H_{D,j'j'}-H_{D,jj}\label{omega:def},\\
\myO_{\al} &=&\bra{N'_\alpha} \Hba \ket {N'_\alpha}-\bra{N_\alpha} H_{B_\alpha} \ket {N_\alpha}=E_{N'_\alpha}- E_{N_\alpha}
\end{eqnarray} 
Eq.~(\ref{comm:S}) follows  immediately from the choice of the jump operators $L_\lambda=\ket{j'}\bra {j}$, while in this Appendix we choose the operators $A_{\al}$ to be eigenoperators of the $\alpha$-th Hamiltonian, with $A_{\al}=\ket{N'_\alpha}\bra{ N_\alpha}$, and $\ket{ N_\alpha}$ the corresponding energy eigenstates.
We also notice that the choice of the interaction Hamiltonian Eq.~(\ref{V:def}) implies that different baths can induce the same transition $j \to j'$, as long as $g_{\alpha, \lambda}\neq 0$. Physically, this corresponds to the case where a bath can be  connected to more than a single subpart of the system, and is therefore more general than the case where one has the same number of baths and subparts, each  bath only inducing transitions in the corresponding subpart of the system. Yet, we deem the approach as ``local", as the eigenkets defining the jump operators are not taken \emph{a priori} to be eigenkets of $H_S$.

We can now study the energy balance for the baths and the system.
First we notice that, since the total Hamiltonian is time-independent, the total energy is always conserved for any $t$ and $\tau$: 

\begin{equation}
\Delta E_{\rm tot}(t,t+\tau)=\tr[ \Ht (\rt(t+\tau) - \rt(t)) ]=0.\label{econs1}
\end{equation} 
If we define
\begin{equation}
\Delta E_x=\average{H_x}_{t+\tau}-\average{H_x}_{t}=\tr[(\E^{\ii \tau \Ht}  H_x \E^{-\ii \tau \Ht}-H_x)\rt(t)],
\label{deltaEx}
\end{equation} 
with $x=B,\, S,V$, energy conservation imposes
\begin{equation}
\Delta E_{B}+\Delta E_{S}=-\Delta E_{V}.
\label{econs2}
\end{equation} 
We have 
\begin{equation}
\E^{\ii \tau \Ht}  H_x \E^{-\ii t \Ht}=H_x+ \ii \tau [\Ht ,H_x]+ \frac{(\ii \tau)^2}{2} [\Ht, [\Ht ,H_x]] + \cdots \label{nestx:eq}
\end{equation} 

Let us first consider $ E_{B}(t+ \tau)- E_{B}(t)$ up to the second order in $\tau$.
From Eq.~(\ref{nestx:eq}) we see that, in order to evaluate the first order contribution to the baths' energy flow, we need to evaluate $[\Ht,H_B]$. To this end, we introduce the operator 
\begin{eqnarray}
 V'_B&=& [\Ht ,H_B]=[V,H_B]\nonumber \\
&=&-\sum_{\al} g_{\al}\myO{\al}(L^\dagger_{ \lambda} A_{\alpha, \lambda} - L_{ \lambda} A^\dagger_{\alpha, \lambda}).
\label{VB:def}
\end{eqnarray} 

To calculate the second order term in  $\Delta E_B$ we need to calculate, 
\begin{equation}
 [\Ht, V'_B]=\left[\sum_\alpha H_{B_\alpha}+H_S+V,V'_B\right],
\end{equation} 
where the individual commutators read 
\begin{widetext}
\begin{eqnarray}
[ H,V'_B]&=& \sum_{\alpha,\lambda} g_{\alpha, \lambda}\myO_{\al}\p{\omega_{\lambda}  (L^\dagger_{ \lambda} A_{\alpha, \lambda} + L_{ \lambda} A^\dagger_{\alpha, \lambda})-([\HND,L^\dagger_{ \lambda}] A_{\alpha, \lambda}-[\HND,L_{ \lambda}] A^\dagger_{\alpha, \lambda}]) },
 \label{comm:HSVB} 
 \\ 
 \pq{\sum_\alpha H_{B_\alpha},\tilde V_B}&=& -\sum_{\alpha,\lambda} g_{\alpha, \lambda}\myO^2_{\al}  (L^\dagger_{ \lambda} A_{\alpha, \lambda} + L_{\lambda} A^\dagger_{\alpha, \lambda}),
 \\ 
\pq{ V,\tilde V_B}&=&-\sum_\alpha\sum_{\lambda,\mu}\pq{g_{\alpha, \lambda}  (L^\dagger_{ \lambda} A_{\alpha, \lambda} + L_{ \lambda} A^\dagger_{\alpha, \lambda}), g_{\alpha, \mu}\myO_{\alpha,\mu}  (L^\dagger_{ \mu} A_{\alpha, \mu} - L_{\mu} A^\dagger_{\alpha, \mu})}, 
\label{V:VB}
\end{eqnarray} 
\end{widetext}
where $\lambda$ and $\mu$ are different transition indices. For future reference, we notice that only the last commutator contains quadratic terms of the types $ A_{\alpha, \lambda}  A^\dagger_{\alpha, \lambda}$ or $ A^\dagger_{\alpha, \lambda}  A_{\alpha, \lambda}$.

We now turn our attention to the system energy balance.
Let us introduce the operators
\begin{eqnarray}
 V'_D &=&{}[V,\HD] =\sum_{\alpha,\lambda} g_{\alpha, \lambda}\omega_{\lambda}(L^\dagger_{ \lambda} A_{\alpha, \lambda} - L_{ \lambda} A^\dagger_{\alpha, \lambda}), \label{VS:def}\\
 V'_{ND} &=&  \pq{V, \HND}\nonumber \\
&=&\sum_{\alpha,\lambda} g_{\alpha, \lambda}({}[L^\dagger_{\lambda},\HND] A_{\alpha, \lambda} +\pq{ L_{ \lambda},\HND} A^\dagger_{\alpha, \lambda}).
\label{VI:def}
\end{eqnarray} 
From Eqs.~(\ref{deltaEx})--(\ref{nestx:eq}), we see that to the first order in $\tau$ we need to evaluate  
\begin{equation}
{}[\Ht,H]=[V,H_S]= V'_D + V'_{ND}.
\end{equation} 
To calculate the second order term in (\ref{nestx:eq}) we need the commutator 
\begin{equation}
[\Ht, V'_D+ V'_{ND}]=[ H_{B}+ \HD+\HND+V,V'_D+ V'_{ND}]. \nonumber
\end{equation} 
We have 
\begin{eqnarray}
{}[H_B, V'_D]&=&\sum_{\lambda,\lambda}g_{\alpha, \lambda} \omega_{ \lambda} \myO_{\al} (L^\dagger_{ \lambda} A_{\alpha, \lambda} + L_{ \lambda} A^\dagger_{\alpha, \lambda}), \nonumber
\\
{}[ H_B,V'_{ND}]&=& -\sum_{\alpha,\lambda} g_{\alpha, \lambda}\myO_{\al}([\HND,L^\dagger_{ \lambda}] A_{\alpha, \lambda}-[\HND,L_{\lambda}] A^\dagger_{\alpha, \lambda}]), \nonumber
 \\
{}[H_S, V'_D]&=&\sum_{\alpha,\lambda}g_{\alpha, \lambda} \omega_{\lambda} \left(-\omega_{ \lambda} (L^\dagger_{ \lambda} A_{\alpha, \lambda} + L_{ \lambda} A^\dagger_{\alpha, \lambda}) \right. \nonumber \\
&&\qquad \qquad \quad \left.  +[\HND,L^\dagger_{ \lambda}] A_{\alpha, \lambda}-[\HND,L_{ \lambda}] A^\dagger_{\alpha, \lambda}] \right), \nonumber
\end{eqnarray} 
\begin{widetext}
\begin{eqnarray} 
{}[V, V'_D]&=& \sum_\alpha\sum_{\lambda,\mu}\pq{g_{\alpha, \lambda}  (L^\dagger_{ \lambda} A_{\alpha, \lambda} + L_{ \lambda} A^\dagger_{\alpha, \lambda}), g_{\alpha, \mu}\omega_{\mu}  (L^\dagger_{ \mu} A_{\alpha, \mu} - L_{ \mu} A^\dagger_{\alpha, \mu})}, \label{V:VS} \\
{}[V, V'_{ND}]&=& \sum_\alpha\sum_{\lambda,\mu}\pq{g_{\alpha, \lambda}  (L^\dagger_{ \lambda} A_{\alpha, \lambda} + L_{ \lambda} A^\dagger_{\alpha, \lambda}), g_{\alpha, \mu}  ({}[L^\dagger_{ \mu},\HND] A_{\alpha, \mu} + {}[L_{ \mu},\HND] A^\dagger_{\alpha, \mu})}.
 \label{V:VI}
\end{eqnarray} 
We notice that the only commutators that are quadratic in the bath operators $A_\lambda$ and $A^\dagger_\lambda$ are the last two.
In accordance with the Born-Markov approximation used to derive the Markovian master equation \cite{Breuer02}, we assume that the total density matrix is factorized at any time, $\rho_{\rm tot}(t)=\rho_S(t)\otimes \rho_B(t)$, that the bath density matrix is time independent: $\rho_B(t)=\bigotimes_{\alpha=1}^{N_b} \rho_{B_\alpha}$, and that $[\rho_{B_\alpha}, H_{B_\alpha}]=0$ even though $\rho_{B_\alpha}$ needs not be an equilibrium state. Without these assumptions the master equation would not be valid or would be modified.

Armed with these assumptions on the baths' state and operators, we can finally calculate the energy flow for the bath and for the system.
From Eqs.~(\ref{deltaEx})-(\ref{V:VB}), we have
\begin{eqnarray}
\Delta E_{B}&=&\ii \tau \average{ V'_B} -\frac {\tau^2}{2} \average{[\Ht, V'_B]} =\nonumber \\
&=&-\tau^2 \sum_{\alpha,\lambda} g^2_{\al}  \myO_{\al} \p{\average{L^\dagger_{ \lambda} L_{ \lambda}}_{S,t} \average{A_{\alpha, \lambda} A^\dagger_{\alpha, \lambda}}_{B}-  \average{L_{ \lambda} L^\dagger_{ \lambda}}_{S,t} \average{A^\dagger_{\alpha, \lambda} A_{\alpha, \lambda}}_{B}}+O(\tau^4),
\label{deltaEB2}
\end{eqnarray} 
where $\langle \bullet \rangle_{S,t} =\tr[{\bullet \rho(t)}]$ and $\langle \bullet \rangle_{B} =\tr[{\bullet \rho_B}]$.
Similarly from Eqs.~(\ref{VS:def})-(\ref{V:VI}) we obtain
\begin{eqnarray}
\Delta E_{S}&=&\tau^2 \sum_{\alpha,\lambda} g^2_{\alpha,\lambda}  \omega_{\lambda} \p{\average{L^\dagger_{ \lambda} L_{ \lambda}}_{S,t} \average{A_{\alpha, \lambda} A^\dagger_{\alpha, \lambda}}_{B}-  \average{L_{ \lambda} L^\dagger_{ \lambda}}_{S,t} \average{A^\dagger_{\alpha, \lambda} A_{\alpha, \lambda}}_{B}}+
\nonumber \\
&&\qquad  + \frac{g^2_{\alpha,\lambda}}{2}\left(\average{ L^\dagger_{ \lambda}[\HND, L_{ \lambda} ] +[L^\dagger_{ \lambda} ,\HND] L_{ \lambda}}_{S,t} \right) \average{A_{\alpha, \lambda} A^\dagger_{\alpha, \lambda}}_{B}+
\nonumber \\
&&  \qquad +  \frac{g^2_{\alpha,\lambda}}{2} \left(\average{ L_{ \lambda}[\HND, L^\dagger_{ \lambda} ] +[L_{ \lambda} ,\HND] L^\dagger_{ \lambda}}_{S,t} \right)\average{A^\dagger_{\alpha, \lambda} A_{\alpha, \lambda}}_{B} +O(\tau^4)\label{deltaES2}.
\end{eqnarray} 
A similar approach for the interaction Hamiltonian leads to
\begin{eqnarray}
\Delta E_{V}&=&-\tau^2 \sum_{\alpha,\lambda}  \frac{g^2_{\alpha,\lambda}}{2}\left[2 (\omega_{\alpha,\lambda} - \myO_{\alpha,\lambda}) \p{\average{L^\dagger_{\alpha, \lambda} L_{\alpha, \lambda}}_{S,t} \average{A_{\alpha, \lambda} A^\dagger_{\alpha, \lambda}}_{B,t}-  \average{L_{\alpha, \lambda} L^\dagger_{\alpha, \lambda}}_{S,t} \average{A^\dagger_{\alpha, \lambda} A_{\alpha, \lambda}}_{B,t}} + \right. \nonumber \\
&&  \qquad\qquad \qquad +\p{\average{ L^\dagger_{ \lambda}[\HND, L_{ \lambda} ] +[L^\dagger_{ \lambda} ,\HND] L_{ \lambda}}_{S,t} } \average{A_{\alpha, \lambda} A^\dagger_{\alpha, \lambda}}_{B}+\nonumber \\
&&  \left. \qquad\qquad \qquad +\p{\average{ L_{ \lambda}[\HND, L^\dagger_{ \lambda} ] +[L_{ \lambda} ,\HND] L^\dagger_{ \lambda}}_{S,t} }\average{A^\dagger_{\alpha, \lambda} A_{\alpha, \lambda}}_{B}\right ] +O(\tau^4)\label{deltaEV2}.
\end{eqnarray} 
The same result can be obtained by invoking energy conservation, Eqs.~(\ref{econs1})-(\ref{econs2}).
\end{widetext}

Comparing Eqs.~(\ref{deltaEB2}) and (\ref{deltaES2}) we see that the amount of energy exchanged by the baths and the system is a linear combination of the $\omega_{\alpha,\lambda}$, i.e. the difference between the diagonal elements of the system Hamiltonian, see Eq.~(\ref{omega:def}). The terms multiplying the energy differences $\omega_{\alpha,\lambda}$ in the first line of Eq.~(\ref{deltaEB2}) or (\ref{deltaES2}) play the role of kinetic terms expressing the rate of such an energy exchange.
The nondiagonal part of the system Hamiltonian does not contribute to heat currents to the baths but is associated with an energy flow between the system and the mechanism connecting the system to the baths, as represented by the Hamiltonian $V$, see Eqs.~(\ref{deltaES2})-(\ref{deltaEV2}). 

As in the main text we define the transition rate for a transition $j \to j'$ and  its reverse
\begin{eqnarray}
\gamma_\alpha(j \to j')&=&\tau g^2_{\alpha,\lambda} \average{A_{\alpha, \lambda} A^\dagger_{\alpha, \lambda}}_{B},
\label{gammajj1:app}\\
\gamma_\alpha(j' \to j)&=&\tau g^2_{\alpha,\lambda} \average{A^\dagger_{\alpha, \lambda} A_{\alpha, \lambda}}_{B}.
\label{gammaj1j:app}
\end{eqnarray} 
Taking the limit $\tau\to 0$ in \eqref{deltaES2}, with $\tau g^2_{\alpha,\lambda}=const.$,
we can  distinguish the two heat currents on the RHS of Eq.~\eqref{deltaES2}
\begin{eqnarray}
\dot Q_{D,\alpha}&=& \sum_{\lambda} \omega_{\lambda} \p{ \gamma_\alpha(j \to j') \average{\pi_j}_{S,t} - \gamma_\alpha(j' \to j) \average{\pi_{j'}}_{S,t} },
\label{Q:D}
\\
\dot Q_{ND,\alpha}&=&\frac{1}{2}\sum_{\lambda}   
\nonumber \\
&&\left[ \gamma_\alpha(j \to j') \p{\average{ L^\dagger_{ \lambda}[\HND, L_{ \lambda} ] +[L^\dagger_{ \lambda} ,\HND] L_{ \lambda}}_{S,t} } + \right .
\nonumber \\
	&+&  \left.   \gamma_\alpha(j' \to j)\p{ \average{ L_{ \lambda}[\HND, L^\dagger_{ \lambda} ] +[L_{ \lambda} ,\HND] L^\dagger_{ \lambda}}_{S,t} }\right ].\nonumber \\ 
\label{Q:ND}
\end{eqnarray} 
where $\pi_j=L^\dagger_\lambda L_\lambda =\ket j \bra j$ as in the main text.
Thus, from Eq.~(\ref{deltaES2}) finally we obtain
\begin{eqnarray}
\dot E_S&=&\sum_\alpha \dot Q_{D,\alpha}+\dot Q_{ND,\alpha}\nonumber \\
&=&\sum_\alpha \tr\p{\rho D^*_\alpha[\HD]+\rho D^*_\alpha[\HND]},
\end{eqnarray} 
which corresponds to the findings of Ref.~\cite{HewgillPRR2021}, and it is the result mentioned in the main text.

Inspection of Eqs.~\eqref{deltaEB2}--\eqref{deltaEV2} can finally help us to address the following question: which are the energy currents that flow from/into the system and the baths? If we assume  the energy gaps in the baths and in the system to be resonant, i.e.
\begin{equation}
\myO_{\al}=\omega_\lambda,
\label{en:bal}
\end{equation} 
a part of the energy currents precisely matches the energy current of the baths: such an energy current is the diagonal heat current \eqref{Q:D} and $\dot Q_{D,\alpha}=-\dot E_{B_\alpha}$ holds. Yet, even if \eqref{en:bal} holds, a part of the energy flows toward the interaction mechanism, represented by the Hamiltonian $V$, and we find $\dot Q_{ND,\alpha}=-\dot E_{V_\alpha}$. Only when the chosen basis $\pg{\ket j}$ is the eigenbasis of $H$ and   \eqref{en:bal} holds, the latter contribution vanishes, and all the energy current flows from the system into the baths.
{Furthermore, from Eq.~\eqref{deltaES2} we also obtain $\dot E_s=\sum_\alpha \dot Q_{D,\alpha}+\dot Q_{ND,\alpha}$. The identical result is obtained by considering $\dot E_S=\average{\dot H}$ and the dynamics in Eq.~\eqref{me:eq}.
Now, in the steady state  $\dot E_S=0$, and thus the RHS of Eq.~\eqref{deltaES2} vanishes. Thus by comparing Eqs.~\eqref{deltaEB2} and \eqref{deltaEV2} we conclude that when the system reaches the steady state the equality $\dot E_{B_\alpha}=-\dot E_{V_\alpha}$ holds too.}

 We notice a difference with the notation of the main text: the interaction Hamiltonian $V$ as given by Eq.~\eqref{V:def} entails both the forward $j \to j'$  jump  ($L_\lambda$) and its inverse  $j' \to j$  ($L^\dagger_\lambda$). Thus the sum over $\lambda$ in Eqs.\eqref{Q:D} entails both the transitions, while in the main text the sum over $\lambda$ in, e.g., Eq.~\eqref{diss}, involves only the transition  $j \to j'$.

It is also worth noticing that, starting from Eq.~(\ref{rhot}) and using the Born-Markov approximation discussed above,  after calculating the difference $\rho(t+\tau)-\rho(t)$ up to the second order in $\tau$, one recovers the local master equation (\ref{me:eq})-- (\ref{diss}) of the main text, with dissipation rates given by Eqs.~(\ref{gammaj1j})--(\ref{gammajj1}). 

\section{Additional information on the numerical results}
\label{num:app}
For each trajectory, the QMC algorithm evolves the system's state from an initial pure state $\ket{\psi(0)}$ to a final state $\ket{\psi(t)}$.
To initiate a QMC trajectory one thus needs to choose an initial state $\ket{\psi(0)}$ compatible with the initial density matrix $\rhoz$. One special choice is to take $\ket{\psi(0)}$ to be one of the $\pg{\ket{k_0}}$, i.e. the eigenstates of $\rhoz$,  with probability $p_{k_0}$, such that:
\begin{equation}
\rhoz=\sum_{k_0} p_{k_0} \ket{k_0}\bra{k_0}.
\end{equation}
 This choice corresponds to a noninvasive measurement that does not induce any back-action on the system.
This is the approach that we followed for the simulations whose results are shown in Fig.~(\ref{figmc1}) in the main text, and in Fig.~\ref{figmc1:app} in this Appendix.
Given that the FT, Eq.~(\ref{FTcoh}), is independent of the initial basis, we also consider below the case where the system is initially projected on the basis $\{\ket {j}\}$ defining the jump operators introduced in Eq.~(\ref{diss}). In contrast to the previous choice, in this case the measurement is indeed invasive and induces measurement back-action. Nonetheless, our fluctuation theorem is still valid.

We now discuss the evaluation of the system's entropy change with the QMC.
We write the quantum FT, Eq.~(\ref{FTcoh}), in the form of its classical counterpart Eq.~(\ref{FTcl}):
\begin{equation}
\average{\E^{-\cS-\SS}}_{\mathrm{tr}}=1,\label{FTtr}
\end{equation} 
where  now $\average{\cdots }_{\mathrm{ tr}}$ is the average over the QMC trajectories.

Given that the generalised entropy $\cS$ along a trajectory is well defined, see Eq.~(\ref{eq:SBsym}) in the main text, we are left with the question of how to sample the system's entropy change $\SS$ along the quantum trajectories.
Inspired by what one does in the classical case one can possibly sample two different quantities along a single trajectory
\begin{eqnarray}
\Delta S_{S,x}&=&-\bra{\psi(t)}\log \rho(t) \ket{\psi(t)}\nonumber 
\\&&+\bra{\psi(0)}\log \rhoz\ket{\psi(0)},\label{eq:sx}\\ 
\Delta S_{S,y}&=&-\log \p{\bra{\psi(t)}\rho(t) \ket{\psi(t)}}\nonumber\\
&&+\log (\bra{\psi(0)} \rhoz\ket{\psi(0)})). \label{eq:sy:app} 
\end{eqnarray} 
where $\rho(t)$ is the solution of the ME (\ref{me:eq}).
In the classical case, these two quantities are equivalent, as $\rho(t)$ is diagonal, and the state $\ket{\psi(t)}$ is one of the states 
of the chosen basis. However, these two possible definitions for the system entropy provide a completely different result for a quantum nondiagonal state. This is also evident in our numerical demonstration of the FT, and we anticipate that the second definition is the correct one.
We can also provide a theoretical argument for this result.
With the QMC, one can evaluate the solution $\rho(t)$ to the ME as 
\begin{equation}
\rho(t)=\average{\ket{\psi(t)}\bra{\psi(t)}}_{\mathrm{ tr}}.
\label{eq:rhotr}
\end{equation} 
Strictly speaking, the equality is exact when one considers the whole ensemble of possible trajectories.
In a QMC trajectory, from $\ket{\psi(0)}=\ket{k_0}$ to $\ket{\psi(t)}$, one can sample $\cS$, Eq.~(\ref{eq:SBsym}). 
Let us now consider the FT in the form of Eq.~(\ref{FTcoh}) and address the question of how we can evaluate $\Psiu(t\,  |\, \pi_{k_0})$ with an average over the MC trajectories.
Inspection of Eq.~(\ref{eq:rhotr}), and of the definition of $\Psi(\xi,t)$, Eq.~(\ref{psi:def}),
suggests
\begin{equation}
\Psiu(t\,  |\, \pi_{k_0})= \average{\left. \E^{-\cS}\right|_{\ket{\psi(t)},\ket{k_0}}\ket{\psi(t)}\bra{\psi(t)}\, }_{\mathrm {tr}}.
\label{psiu:num}
\end{equation} 
Comparison of the last equality with the FT in the form of Eqs.~(\ref{FTcoh}) and ~(\ref{FTtr}), suggests that  $\Delta S(t)_{S,y}$ as given by  Eq.~(\ref{eq:sy:app}) must be used in the numerical evaluation of the FT.

This conclusion is confirmed by the results for the two-spin system shown in Fig.~\ref{figmc1:app} where the trajectory average in Eq.~(\ref{FTtr})  is shown for different values of the system parameters and of the initial state $\rhoz$, and the two possible definitions of system entropy change (\ref{eq:sx})--(\ref{eq:sy:app}) are used. In particular, in Fig.~\ref{figmc1:app}-(d) we consider the case of a time-dependent Hamiltonian, where the external field is changed according to $h(t)=h_0+(h_1-h_0) t/\tf$. While panels (a)-(d) in Fig.~\ref{figmc1:app} correspond to the case where the baths are thermal, with dissipation rates obeying the  LDBC, in panel (e) we consider the case where the dissipation rates violate the detailed balance, see below for further details.

\begin{figure}[h]
\center
\includegraphics[width=8cm]{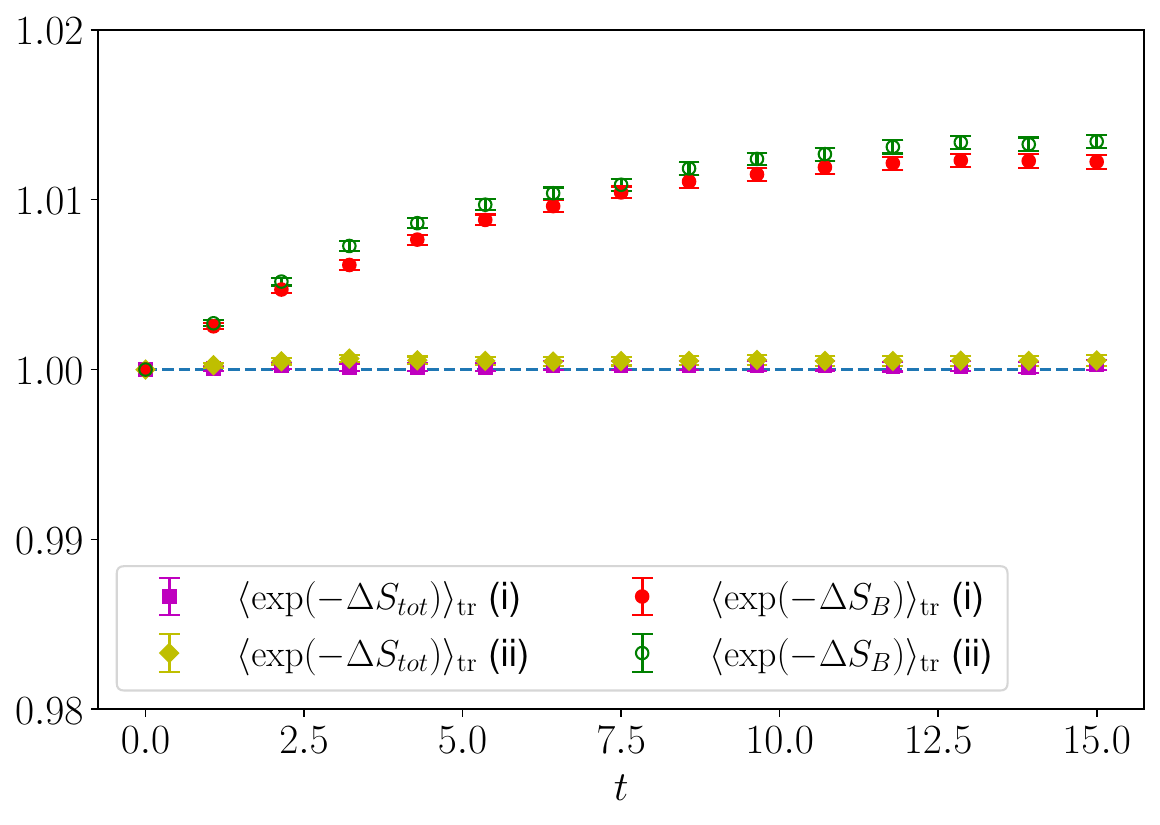}
	\caption{QMC simulations of the two-spin system (\ref{H:spins}): $\average{\exp\p{-\Delta S_{\rm tot}}}_{\mathrm{tr}}$ and $\average{\exp\p{-\SB}}_{\mathrm{tr}}$  as functions of $t$, with $\Delta S_{\rm tot}=\SB+\SS$.    Nondiagonal $\rhoz$, $J=0.1$, $h=0.2$,  $T_b=1.2$.  See Appendix \ref{other:app} for the detailed description of $\rhoz$. We start the simulations with (i) the basis $\{ \ket{k_0}\}$ that diagonalises the initial state $\rhoz$, and (ii) the basis $\{\ket j\}$ defining the jump operators introduced in (\ref{diss}). The dissipation  rates \eqref{diss_rat} have been used.}
\label{fig:in:basis}
\end{figure}

We can also check that the FT holds for an arbitrary initial basis, as predicted by Eq.~(\ref{FTcoh}). In Fig.~\ref{fig:in:basis} we start the simulations from (i) the basis $\{\ket{k_0}\}$ that diagonalises the initial state $\rhoz$, and (ii) the basis $\{\ket j\}$ defining the jump operators introduced in Eq.~(\ref{diss}).
We see that the FT is numerically satisfied for both choices.

\subsection{Other information on the numeric results}
\label{other:app}

We remind the reader that in  the main text we consider  a system of two spin-1/2 particles with Hamiltonian 
\begin{equation}
H=-J \sigma_{x,a}\otimes \sigma_{x,b}-h( \sigma_{z,a}\otimes \II_{2,b}+\otimes \II_{2,a} \otimes \sigma_{z,b})
\label{H:spins}
\end{equation} 
where each spin is connected to an equilibrium reservoir at temperatures $T_a$ and $T_b$, respectively. The jump operators ``flip'' the individual spins: $L_\lambda=\sigma_{-,a}\otimes \II_b$ or $L_\lambda=\II_a \otimes \sigma_{-,b}$. 

\begin{figure}
\center
\parbox{.32\textwidth}{\includegraphics[width=.32\textwidth]{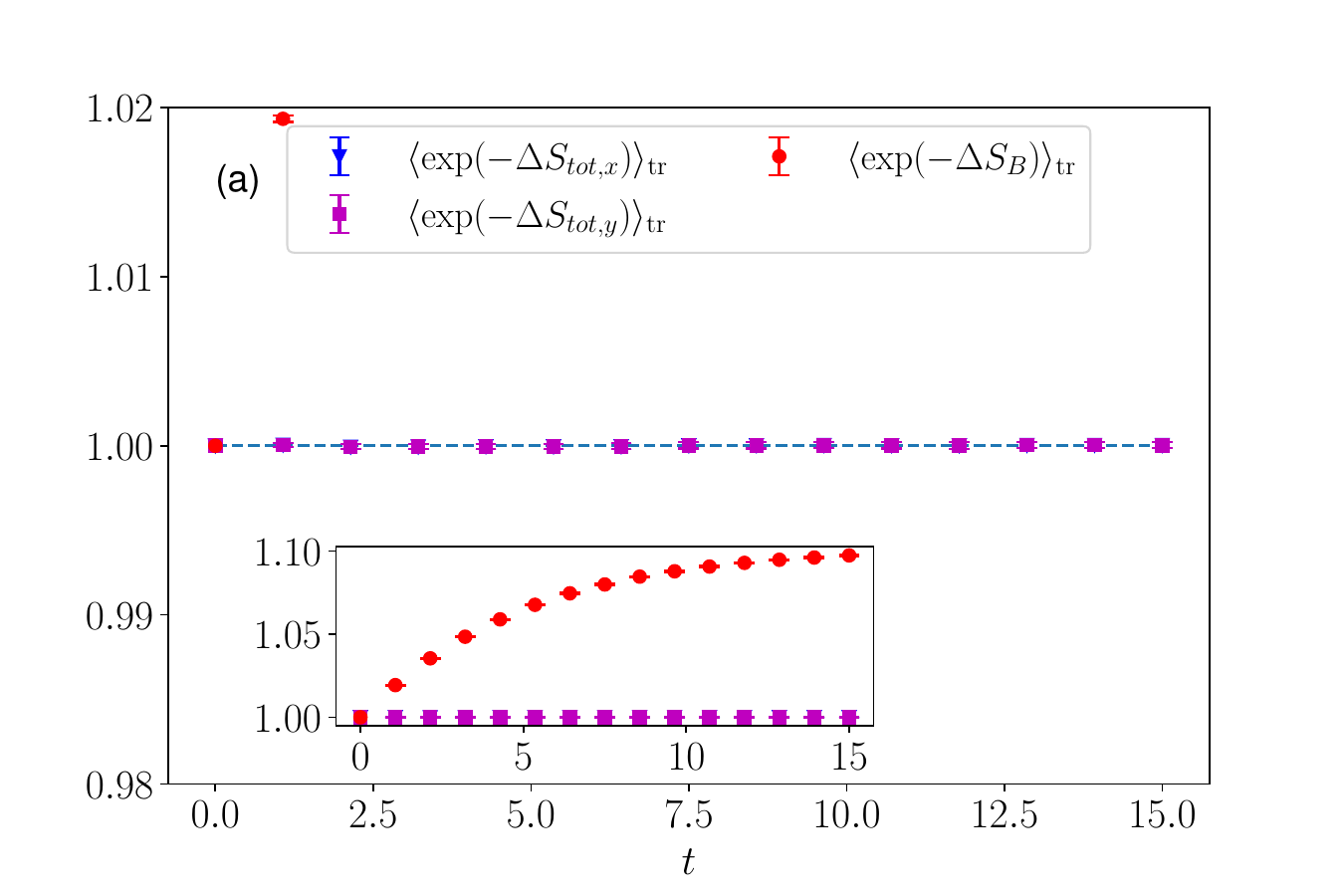}}
\hfill
\parbox{.32\textwidth}{\includegraphics[width=.32\textwidth]{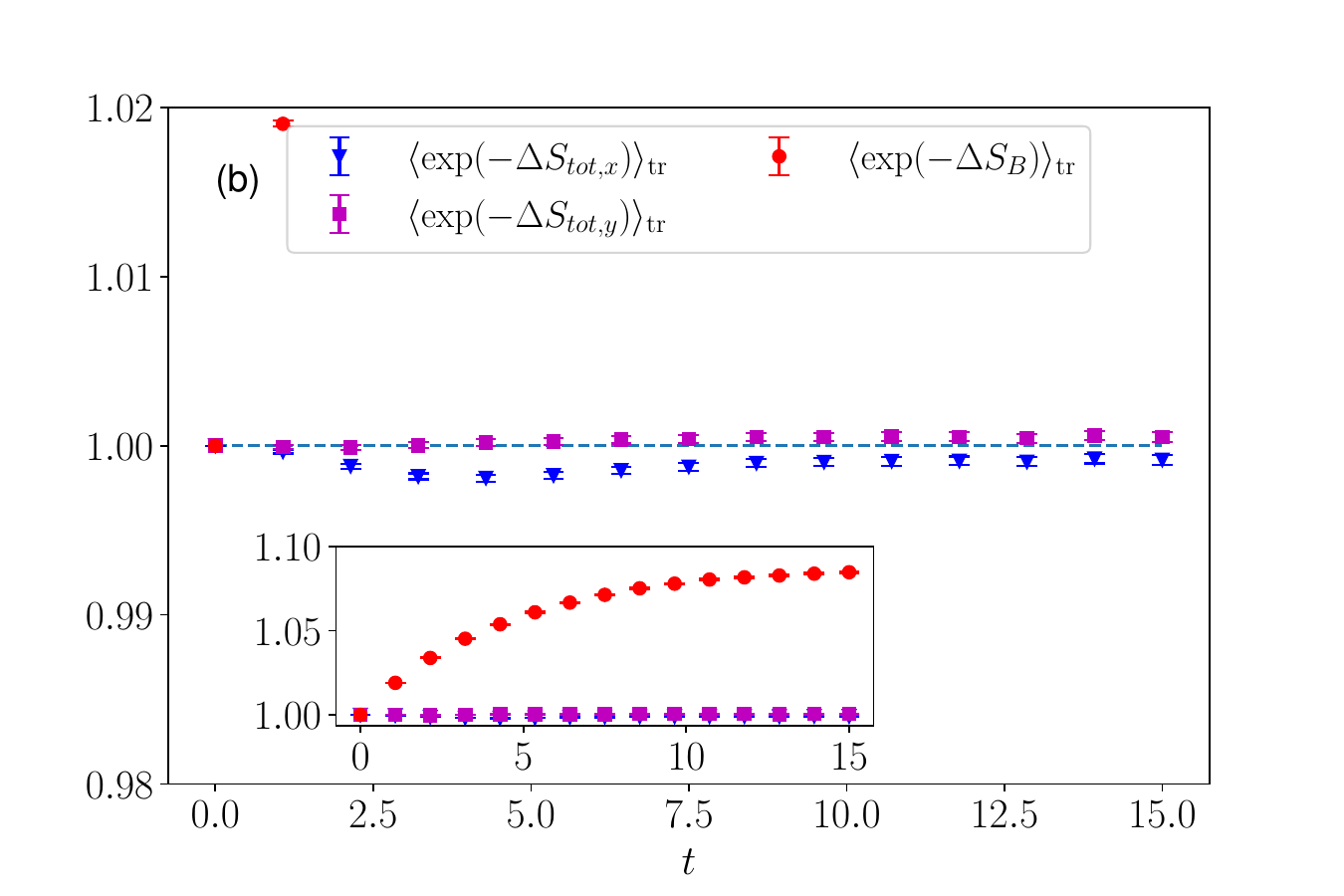}}
\hfill
\parbox{.32\textwidth}{\includegraphics[width=.32\textwidth]{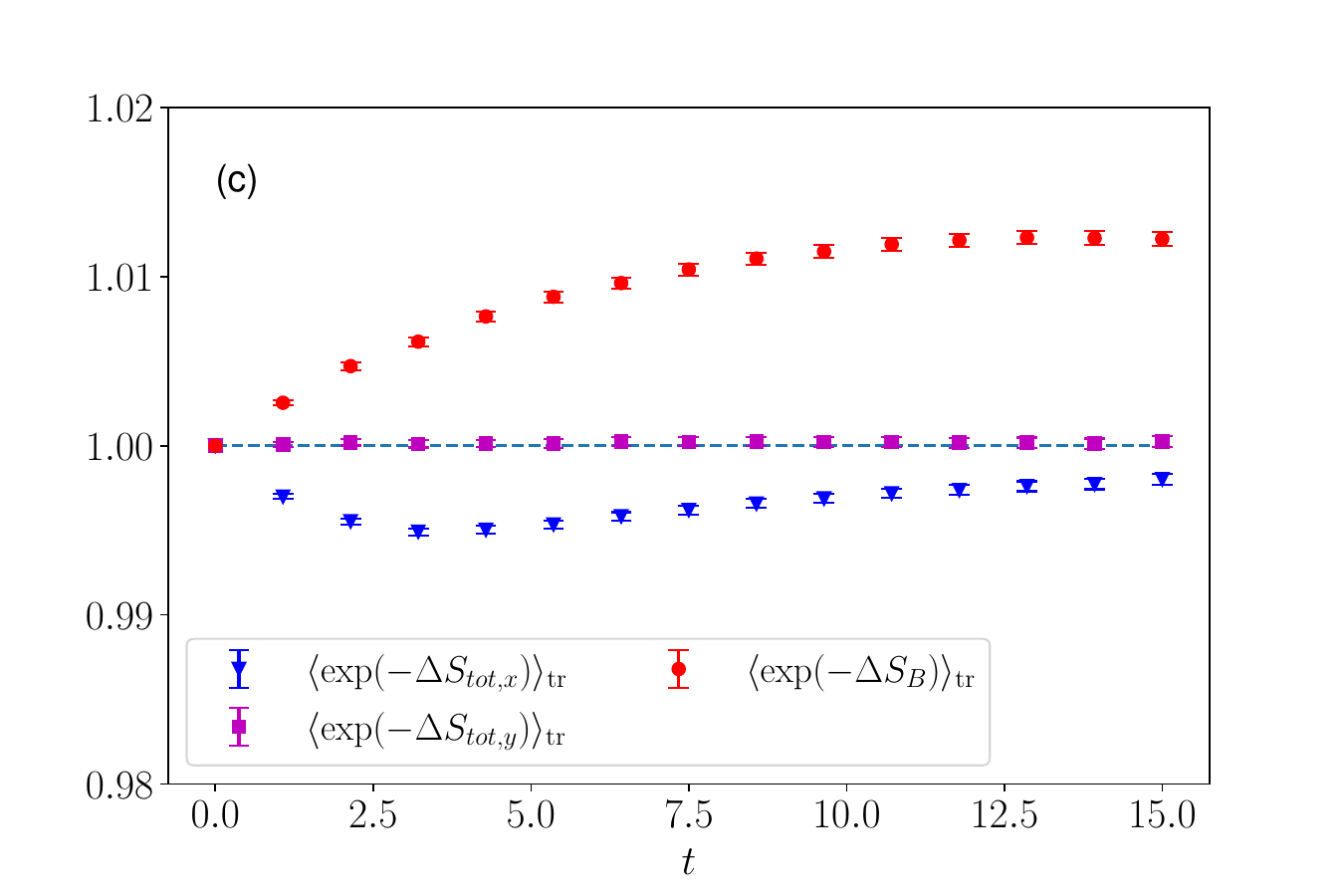}}

\bigskip
\flushleft
\parbox{.32\textwidth}{\includegraphics[width=.32\textwidth]{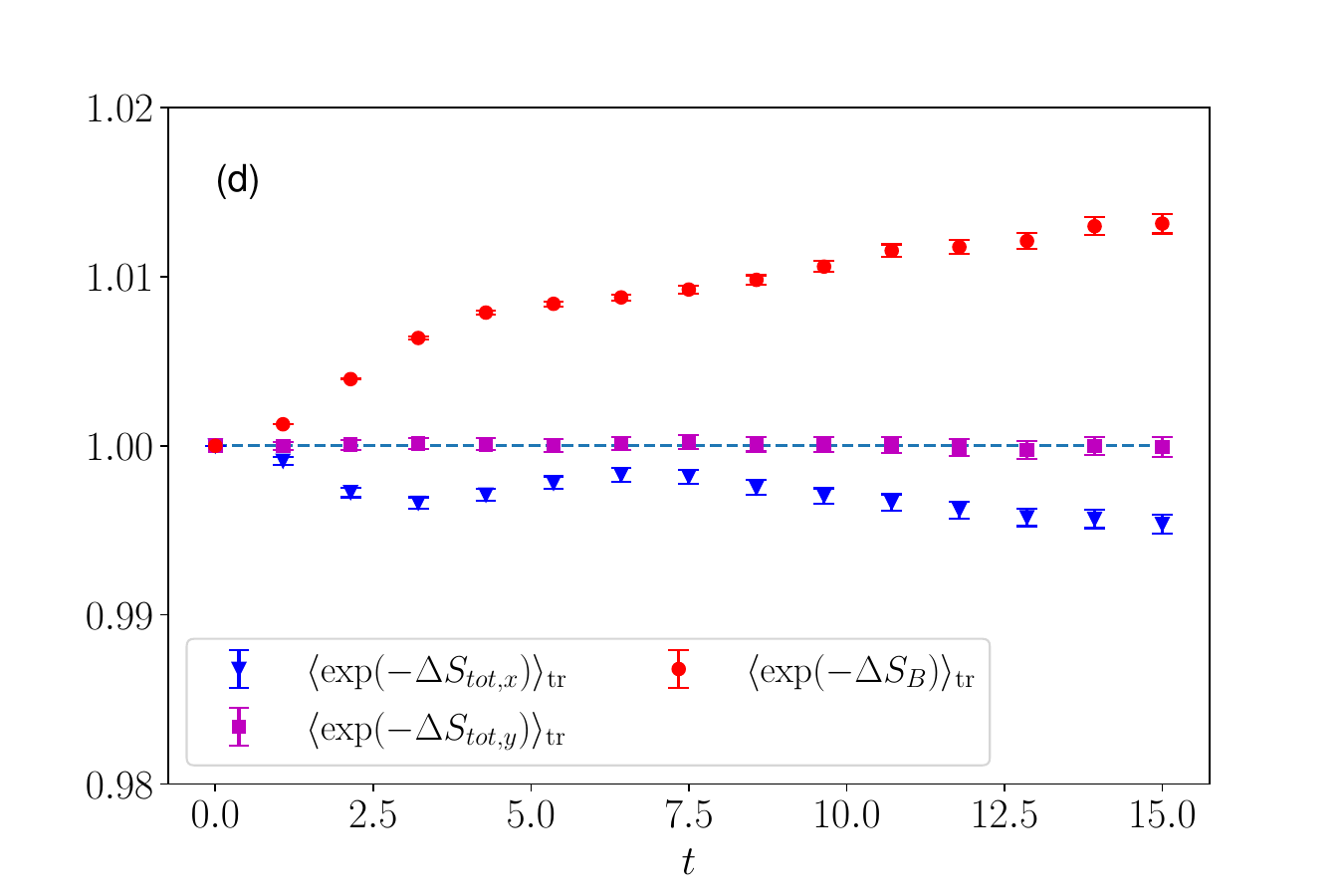}}
\parbox{.32\textwidth}{\includegraphics[width=.28\textwidth]{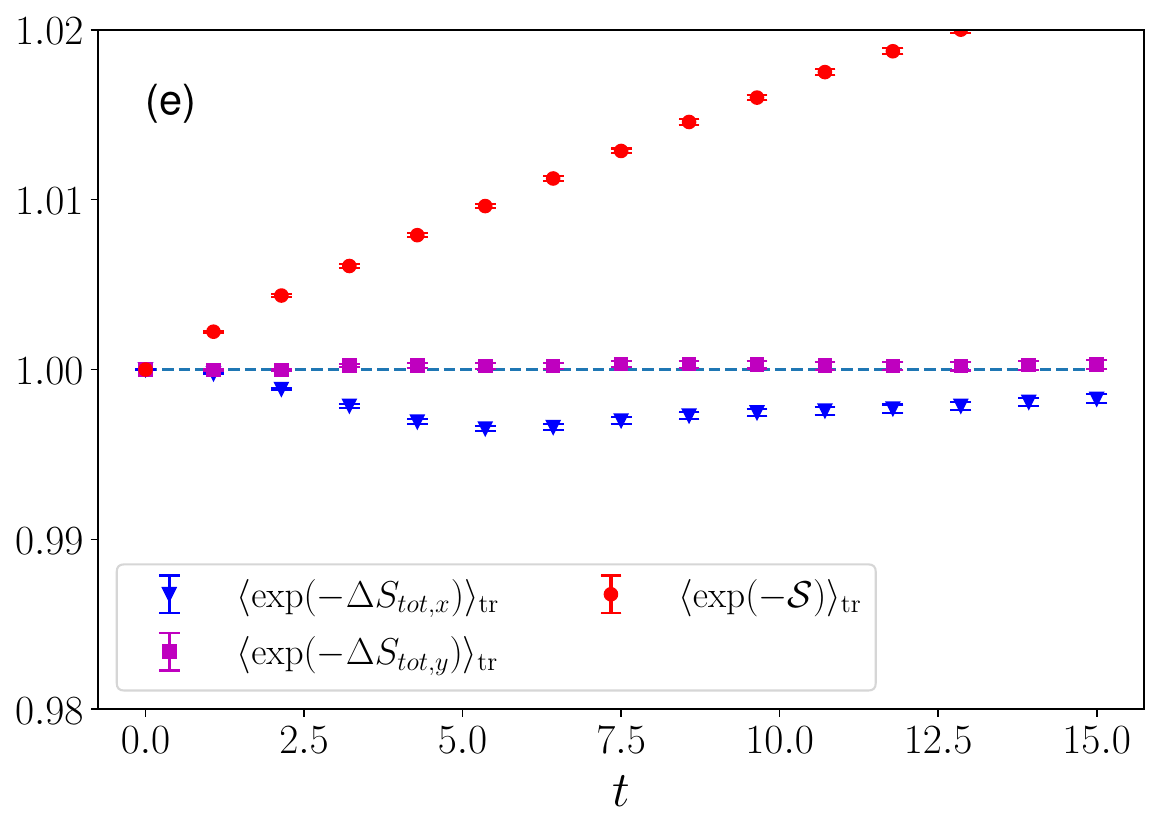}}
\caption{QMC simulations of the two-spin system (\ref{H:spins}): $\average{\exp\p{-\Delta S_{\rm tot}}}_{\mathrm{tr}}$ and $\average{\exp\p{-\SB}}_{\mathrm{tr}}$  as functions of $t$, with  $\Delta S_{\rm tot}=\cS+\SS$. For the system's entropy variation we use the two possible definitions Eq.~(\ref{eq:sx}) and (\ref{eq:sy:app}).  Averages over $10^6$ trajectories, $g=0.1$, $T_a=1$.  (a): diagonal $\rhoz$, $J=0$, $h=0.2$,  $T_b=1$ (classical case at equilibrium). (b):  diagonal $\rhoz$, $J=0.1$, $h=0.2$,  $T_b=1.2$. (c):  nondiagonal $\rhoz$, $J=0.1$, $h=0.2$,  $T_b=1.2$. (d): nondiagonal $\rhoz$, $J=0.2$,  $T_b=1.2$, and time dependent field $h(t)=0.4 t/\tf$, with $\tf=15$. See section \ref{other:app} for the detailed description of $\rhoz$. In panels (a)-(d) the jump rates \eqref{diss_rat} are used, which satisfy the LDBC, and thus $\cS=\Delta S_B$. (e) Non-thermal jump rates $\gamma_{\alpha,\lambda}$, as reported in table \ref{table1}, that do not satisfy the LDBC, with diagonal $\rhoz$, and $J=0.1$, $h=0.2$. }
\label{figmc1:app}
\end{figure}

The diagonal initial state for Fig.~(\ref{figmc1}) in the main text and  Fig.~\ref{figmc1:app} in this Appendix (panels (a), (b), and (d)) reads
$\rhoz_D=(0.4,0.275,0.175,0.15)$ in the basis $ \pg{\ket{\uparrow \uparrow  },\ket{\uparrow \downarrow  }, \ket{\downarrow\uparrow   }, \ket{\downarrow \downarrow  }}$ .

The nondiagonal initial state in the same figures (panel (c)) was obtained as follows.
We introduced the operator
\begin{eqnarray}
H_0&=&-\frac 1 2  \pi_{\ket{\uparrow \uparrow  }} -\frac 1 4 \pi_{\ket{\downarrow \downarrow  }} -\frac 1 3  \pi_{\ket{\uparrow \downarrow  }}\nonumber \\
&& -h\sigma_{x,a} \otimes \II_2+ h\II_2 \otimes \sigma_{x,b}, 
\end{eqnarray} 
then 
\begin{equation}
\rhoz=\E^{-H_0/2}/\tr[\E^{-H_0/2}].
\end{equation} 

\onecolumngrid

\twocolumngrid

\begin{table}[h]
\begin{tabular}{rl}
	$\gamma_{a} (0\to 2)$= & 0.0813  \\
	$\gamma_{a} (2\to 0)$= & 0.1335  \\
	$\gamma_{a} (1\to 3)$= & 0.0976  \\
	$\gamma_{a} (3\to 1)$= & 0.1407  \\
$\gamma_{a} (j\to j)$=& 0.1  \\
        $\gamma_{b} (0\to 1)$=& 0.1011  \\
	$\gamma_{b} (1\to 0)$= & 0.1270  \\
	$\gamma_{b} (2\to 3)$= & 0.1112  \\
	$\gamma_{b} (3\to 2)$= & 0.1411 \\ 
	$\gamma_{b} (j\to j)$= & 0.12  
	\end{tabular}
	\caption{Dissipation rates for the different transitions without the LDBC. We enumerate the standard basis as follows $\pg{\ket 0, \ket 1, \ket 2, \ket 3}=\pg{\ket{\uparrow \uparrow  },\ket{\uparrow \downarrow  }, \ket{\downarrow\uparrow   }, \ket{\downarrow \downarrow  }}$ }
\label{table1}
\end{table}

For the case where the LDBC holds, 
the  bosonic bath dissipation rates used to solve numerically the quantum ME (\ref{me:eq}) and in the QMC algorithm read \cite{Breuer02}
\begin{align}
&\gamma_{\alpha,\lambda}=\dfrac{g|\omega_{\lambda}|}{1-e^{\beta_\alpha|\omega_{\lambda}|}}
\begin{cases} 
e^{-\beta_\alpha \omega_{\lambda}} & \omega_{\lambda}\geq 0 ,\\
 1 & \omega_{\lambda}\leq 0,
\end{cases}
\label{diss_rat}
\end{align}
where $g$ is a microscopic frequency and $\omega_\lambda$ is defined in Eq.~(\ref{omega:def}). 
When the LDBC holds, the generalized entropy $\cS$, Eq.~(\ref{eq:SBsym}) takes the simple form  
\begin{eqnarray}
\cS&=&\SB=-(\beta_a Q_{a,D}+\beta_b Q_{b,D})\nonumber \\
&&=-\sum_{l=1}^{n_a}\beta_a \omega_{j_{l+1},j_l
}(t_l)-\sum_{m=1}^{n_b}\beta_b \omega_{j_{m+1},j_m}(t_m),
\label{app:SBsym}
\end{eqnarray} 
where $\SB$ is the entropy change in the bath.

For the case where the LDBC does not hold (panel (e) in fig.~\ref{figmc1:app} in this Appendix, and in fig.~\ref{figmc1} in the main text), we choose  the dissipation rates as in table \ref{table1}.

\section{Complement to Section \ref{ME:subsec}}
\label{unitary:ME:app}
This Appendix complements the derivation of the results in Sec.~\ref{ME:subsec}.

We first prove Eq.~\eqref{comm:tilde}.
We start by introducing the non-Hermitian  operator
\begin{equation}
V'_{\al}= g_{\al} (L^\dagger_\lambda  A_{\al}- L_\lambda A^\dagger_{\al}), 
\label{V1:alpha}
\end{equation}
and notice that
\begin{eqnarray}
\comm {\Hb}{\Ht}&=&\comm {\Hb}{V}=\sum_{\al} \Omega_{\al} V'_{\al} ,\label{comm1}\\
\comm {\Hb}{\comm {\Hb}{\Ht}}&=&[ \Hb, \sum_{\al}  \Omega_{\al} V'_{\al}]\nonumber \\
& =&\sum_{\al} \Omega^2_{\al} V_{\al}. \label{comm2}
\end{eqnarray}

We proceed by noticing that for two noncommuting operators $X$ and $Y$, the following equality holds
\begin{equation}
\E^X \E^Y=\E^{Y + \comm X Y + \frac 1{2!} [X,[X,Y]]+  \frac 1{3!} [X,[X,[X,Y]]]+ \cdots }\,  \E^X.
\end{equation} 
We now turn our attention back to Eq.~\eqref{Psrev}, and taking $X=-S_B/2$ and $Y=\Ht$ in the previous equation, we find 
\begin{eqnarray}
&&\E^{-\frac{S_B} 2}\E^{-\ii t \Ht}\nonumber \\
&=&\E^{-\ii t\pq{\Hs + \Hb + \sum_{\al} \sum_{n=0}^\infty g_{\al} \p{ y_{\al}^{2 n} \frac{V_{\al}}{2 n!}- y_{\al} ^{2 n+1} \frac{V'_{\al}}{(2 n+1)!}  }}}\E^{-\beta_\alpha \Hb/2}\nonumber \\
&=&\E^{-\ii t\pq{\Hs + \Hb + \sum_{\al}  \bar V_{\al}}} \E^{-\frac{S_B} 2}
\label{comm:tilde:app}
\end{eqnarray} 
where $y_{\al}=\beta_\alpha \Omega_{\al}/2$  and 
\begin{equation}
\bar V_{\al}=g_{\al} \p{ \E^{-\beta_\alpha \Omega_{\al}/2} L^\dagger_\lambda A_{\al} + \E^{\beta_\alpha \Omega_{\al}/2} L_\lambda A^\dagger_{\al}   }
\label{Vtilde:app}
\end{equation} 
is a modified, non-Hermitian interaction Hamiltonian that enters in Eq.~\eqref{Vtilde}.

We now turn our attention to Eq.~(\ref{rhost}), and show how it is equivalent to Eq.~\eqref{gen:ME:un} and thus to Eq.~\eqref{eqPsidual} in the limit $\tau \to 0$.  
By using the definition for the rates $\gamma_\alpha(j\to j')=  \gamma_{\alpha, j'j}$ introduced in Eqs.~\eqref{gammajj1}--\eqref{gammaj1j}, Eq.~(\ref{rhost}) becomes 
\begin{eqnarray}
&&\frac{ \Psi(t+\tau)- \Psi(t)} \tau \simeq -\ii [\Hs,\Psi(t)]\nonumber \\
&&+ \sum_{\al} \gamma_{\alpha, j'j}  \E^{\beta_\alpha \Omega_{\al}}  L_\lambda \Psi(t) L_\lambda^\dagger + \sum_{\al} \gamma_{\alpha,j j'}  \E^{-\beta_\alpha \Omega_{\al}} L_\lambda^\dagger \Psi(t) L_\lambda  \nonumber \\
&& -\frac{1} {2}  \sum_{\al}   \gamma_{\alpha,j'j} \pg{  L_\lambda^\dagger  L_\lambda, \Psi(t) }  +  \gamma_{\alpha,j j'} \pg{  L_\lambda  L_\lambda^\dagger, \Psi(t) } \nonumber\\
&=& -\ii  [\Hs,\Psi(t)]\nonumber \\
&& + \sum_{\al} \gamma_{\alpha, j j'}    L_\lambda \Psi(t) L_\lambda^\dagger +\gamma_{\alpha,j' j}   L_\lambda^\dagger \Psi(t) L_\lambda  \nonumber \\
&& -\frac{1} {2}  \sum_{\al}   \gamma_{\alpha,j' j} \pg{  L_\lambda^\dagger  L_\lambda, \Psi(t) }  +  \gamma_{\alpha,j j'} \pg{  L_\lambda  L_\lambda^\dagger, \Psi(t) }. 
\label{rhost1}
\end{eqnarray} 
where in the last equality we have used Eq.~(\ref{dbB:eq}).
A careful inspection  shows that the diagonal part of the RHS of Eq.~\eqref{dbB:eq} corresponds to Eq.~\eqref{PsiD1:app}, while the nondiagonal part of~(\ref{dbB:eq}) corresponds to Eq.~\eqref{PsiND1:app}.
This proves that the nonunitary dynamics introduced in Eq.~\eqref{Psrev:tilde1} leads to the modified ME~\eqref{eqPsidual} under the same approximations that lead to the GKSL ME~\eqref{me:eq}.

\bibliography{bibliography}

\end{document}